\documentclass{article}
\usepackage{PRIMEarxiv}
\usepackage{graphicx} 
\usepackage{subcaption}
\usepackage[font=small,skip=0pt]{caption} 
\usepackage{color}
\usepackage{xcolor}
\usepackage{xspace}
\usepackage{amsmath}
\usepackage{fancyhdr}
\usepackage{hyperref}
\usepackage{amsfonts}

\newcommand{\p}{\textit{Producer}\xspace}
\newcommand{\w}{\textit{Worker}\xspace}
\newcommand{\co}{\textit{Consumer}\xspace}
\newcommand{\slowd}{\textbf{WF}\xspace}
\newcommand{\slocdf}{\textbf{CF}\xspace}
\newcommand{\slocdr}{\textbf{CR}\xspace}
\newcommand{\slowc}{\textbf{W-consumption}\xspace}
\newcommand{\sloi}{\textbf{Latency}\xspace}
\newcommand{\slocc}{\textbf{C-consumption}\xspace}
\newcommand{\slos}{\textbf{Success}\xspace}
\newcommand{\slod}{\textbf{Smoothness}\xspace}
\newcommand{\actInc}{\emph{Increase}\xspace}
\newcommand{\actDec}{\emph{Decrease}\xspace}
\newcommand{\actSt}{\emph{Stay}\xspace}
\newcommand{\actOff}{\emph{Switch off}\xspace}
\newcommand{\actOn}{\emph{Switch on}\xspace}
\newcommand{\actEn}{\emph{Enable}\xspace}
\newcommand{\actDis}{\emph{Disable}\xspace}

\title{
Distributed Intelligence in the Computing Continuum with Active Inference 
}

\author{
  Victor Casamayor Pujol \\
  Universitat Pompeu Fabra \\
  Barcelona, Spain \\
  \texttt{victor.casamayor@upf.edu} \\
  \And
  Boris Sedlak \\
  TU Wien, Distributed Systems Group \\
  Vienna, Austria \\
  \And
  Tommaso Salvatori \\
  VERSES AI \\
  Los Angeles, USA \\
  \And
  Karl Friston \\
  University College London \\
  London, UK \\
  \And
  Schahram Dustdar \\
  TU Wien, Distributed Systems Group Vienna, Austria \\
  Universitat Pompeu Fabra Barcelona, Spain \\
}

\date{May 2025}

\begin{document}

\maketitle

\begin{abstract}
The Computing Continuum (CC) is an emerging Internet-based computing paradigm that spans from local Internet of Things (IoT) sensors and constrained edge devices to large-scale cloud data centers. Its goal is to orchestrate a vast array of diverse and distributed computing resources to support the next generation of Internet-based applications. However, the distributed, heterogeneous, and dynamic nature of CC platforms demands distributed intelligence for adaptive and resilient service management.
This article introduces a distributed stream processing pipeline as a CC use case, where each service is managed by an Active Inference (AIF) agent. These agents collaborate to fulfill service needs specified by \emph{SLOiDs}, a term we introduce to denote Service Level Objectives that are aware of its deployed devices, meaning that non-functional requirements must consider the characteristics of the hosting device.
We demonstrate how AIF agents can be modeled and deployed alongside distributed services to manage them autonomously. Our experiments show that AIF agents achieve over 90\% SLOiD fulfillment when using tested transition models, and around 80\% when learning the models during deployment. We compare their performance to a multi-agent reinforcement learning (MARL) algorithm, finding that while both approaches yield similar results, MARL requires extensive training, whereas AIF agents can operate effectively from the start.
Additionally, we evaluate the behavior of AIF agents in offloading scenarios, observing a strong capacity for adaptation. Finally, we outline key research directions to advance AIF integration in CC platforms.
\end{abstract}

\section{Introduction}
The next generation of Internet-based applications is poised to change how we live as a society: Autonomous driving will revolutionize urban transportation, demanding ultra-low latency to enable real-time decision-making~\cite{lin_architectural_2018}. E-Health applications will allow for remote and detailed patient care, but will also require local data processing to protect patient privacy~\cite{makina_survey_2024}. Fleets of robots will autonomously clean and maintain city streets, ensuring efficient urban management~\cite{zahidi_optimising_2024}. Resource usage, such as electricity and water~\cite{shahra_intelligent_2024}, will be optimized through advanced distribution systems, and AR/VR technologies will reshape how we interact with people and objects, necessitating rapid processing of large data streams~\cite{sukhmani_edge_2019}.
All these applications share the need for near-real-time computations while processing large amounts of data, which the Cloud alone cannot provide due to transmission latency~\cite{hu_edge_2020}.

The Computing Continuum (CC) is considered by the community as the emerging platform with the potential capabilities to bring the required performance to these new applications \cite{beckman_harnessing_2020, dustdar_distributed_2023,nardelli_compute_2024}.
Integral to this continuum are its inherent heterogeneity and dynamism, fundamental aspects that must be addressed when developing effective management methods~\cite{casamayor_pujol_fundamental_2023}. These characteristics impact multiple facets of the system. For instance, heterogeneity is present in the types of computing and networking units, data modalities, system providers, and the variety of services within applications. Device heterogeneity, for example, hinders centralized auto-scalers because adaptations must be tailored to each device’s capacity and characteristics~\cite{zhang_heteroedge_2019}. Similarly, managing multiple system providers requires agreements and methodologies that go beyond current Cloud solutions, which typically rely on a single provider handling the entire infrastructure~\cite{casamayor_pujol_deepslos_2024}.
Dynamism, on the other hand, affects devices, networking capacities, user demands, and system costs~\cite{baek_three_2020}. Devices may have fluctuating capacities, making fast re-deployment or computation offloading necessary~\cite{hu_cec_2023}. Furthermore, dynamic costs and the involvement of multiple stakeholders require transparent rules to ensure fair and efficient application deployment and execution.

As a unified computing fabric that integrates all current computational tiers (i.e., IoT, Edge, Fog, and Cloud), the CC paradigm holds great potential to meet application requirements by harnessing the strengths of each tier~\cite{beckman_harnessing_2020}. The CC achieves this by placing the services’ elasticity strategies directly where they are most needed, i.e., where services execute. Elasticity strategies refer to the ability of software services to adjust their configuration in response to changing conditions to maintain performance~\cite{furst_elastic_2018}. In Cloud environments, a common elasticity strategy is scaling the number of service replicas. In the Computing Continuum, however, strategies like offloading services to other devices or adapting the quality of data for processing are more common and effective~\cite{manaouil_kubernetes_2020}.
Local adaptation of services becomes essential to ensure the necessary performance for these applications, given the inherent heterogeneity and dynamism of the CC. This shift presents an exciting opportunity to move beyond the current Cloud-inspired model, paving context sensitivity for more advanced adaptation techniques.

Local adaptation implies that services must be able to monitor their status —what we refer to as Service Level Objectives in Device (SLOiD)— as well as their surrounding environment, which includes other related services, the hosting device, and even user behavior patterns. They need to analyze this information and then ask themselves: \textit{Is an elasticity strategy necessary to improve current SLOiD fulfillment?} If so, \textit{which one?} Based on this analysis, services must then adapt in real-time to meet current needs.
This process requires services to continuously collect and interpret data to infer both the system’s and the environment's state. These real-time decisions are crucial for maintaining performance but pose a complex planning challenge, as they must consider multiple factors and adjust actions on-the-fly. To address this challenge, we leverage Active Inference (AIF)~\cite{friston_active_2016}, a probabilistic agent-based approach that enables processing services to continuously adapt to a dynamic environment. From a technical perspective, AIF is particularly apt for the multiple and federated constraint problem furnished by SLOiDs; AIF inherits from the free energy principle, under which self-organisation is specified in terms of minimising surprisal or self-information. In AIF, this surprise is specified in terms of the negative log probability of occupying a preferred or characteristic state. Crucially, this means the objective function in AIF are functionals of probability distributions over outcomes — as opposed to simple functions of outcomes per se, such as reward, utility or cost-to-go. This means that the specification of constraints is naturally accommodated in terms of a probability distribution over all states or outcomes a system can experience. Furthermore, because free energy is an extensive quantity, if all the agents in a distributed architecture minimise their (expected) free energy, then the joint free energy of the ensemble is also minimised. This licences local (planning as) inference. In what follows, we leverage these two fundaments of AIF in the context of CC.

This article presents an AIF multi-agent system for managing a pipeline of distributed services. Each service agent is modeled as a partially observable Markov decision process (POMDP), which continuously evaluates its state and selects self-adaptive actions in real-time to optimize performance. To the best of our knowledge, this is the first attempt to use AIF and POMDPs in distributed processing pipelines of the CC. Hence, we present a detailed description on how we leverage AIF to encourage others researchers to test this novel approach.
In addition, we discuss a methodology for integrating these agents into CC applications and provide a research roadmap outlining the requisite developments to make this technology fully actionable for real-world environments.

The rest of this article is organized as follows. We present the background in Section~\ref{sec:background} needed to follow the article, specifically we discuss SLOiDs, POMDPs and AIF. Then, in Section~\ref{sec:vision} we introduce the overall vision for application management in the CC based on the key idea of distributing intelligence throughout the platform. Section~\ref{sec:methodology} presents the methodology used to leverage AIF agents for a CC use case. The results of the experiments are presented in Section~\ref{sec:results}, which are used in Section~\ref{sec:discussion} as a starting point to describe future developments that need to be addressed to use AIF in the CC. Afterwards, in Section~\ref{sec:rw} we present related work and summarize our findings in Section~\ref{sec:conclusions}.

\section{Background}
\label{sec:background}
In this section, we discuss the three concepts that are fundamental to the contributions made in this paper. First, we present SLOiDs as prevalent mechanisms for ensuring high-level service requirements. To ensure SLOiDs, we highlight how AIF, an emerging framework from neuroscience, can support the necessary resilient decision-making. In more detail, we introduce how these AIF-driven decision-making processes are modeled and optimized during runtime using POMDPs. 

\subsection{Service Level Objectives in Devices -- SLOiDs}
In Cloud computing, the promised service quality between infrastructure provider and application developer is specified within a Service Level Agreement (SLA). There, both entities agree on specific service-related measurements (Service Level Indicators - SLIs) that must behave in a specified manner (Service Level Objectives - SLOs), e.g., the CPU utilization of a computing node must stay under 80\%~\cite{beltran_defining_2016}. Then, the number of \textit{good events}, e.g., number of measurements where the CPU stayed under 80\%, is divided by the number of \textit{valid events}, e.g., number of measurements considered, for a determined period of time~\cite{beyer_site_2016}. 
For a Cloud business model, this value provides a notion of reliability of the service, which is used to define penalties (e.g., monetary compensations) for the infrastructure provider in case SLOs are violated. Indeed, the infrastructure provider will use elasticity strategies, e.g., scaling system resources up, when the SLO is not being fulfilled to reestablish the system's desired behavior~\cite{dustdar_principles_2011}.

\vspace{5pt}
\textbf{SLO Fulfillment:} A probabilistic estimate of the system's performance which we aim to optimize. It is defined as the fraction of \textit{good events} divided by the \textit{valid events}, computed over a given period of time.
\vspace{5pt}

Applications distributed over the CC pose a variety of requirements (e.g., latency and/or quality) that define how each device and component should operate. However, contrarily to traditional Cloud computing, enforcing SLOs in the CC requires orchestration over numerous heterogeneous devices~\cite{nardelli_compute_2024}. To capture the complexity this adds, we extend the SLO definition and call them Service Level Objectives in Devices. 
In that regard, the key aspects that characterize SLOiDs, also in relation to regular SLOs, are as follows:

\begin{itemize}
    \item Hardware heterogeneity directly dictates SLOiDs feasibility. Unlike the abstracted and relatively homogeneous resources in traditional Cloud environments, the CC is characterized by hardware diversity. SLOiDs cannot be practically defined without considering the specific capabilities and limitations (e.g., CPU power, memory availability, energy constraints, network bandwidth) of the target deployment devices. For instance, a latency SLO (e.g., response time below 50ms) achievable on a powerful Edge server is impracticable on a resource-constrained IoT sensor. Therefore, SLOiDs must be tailored to specific device classes. Ignoring this leads to perpetually unachievable objectives on lower-end devices or over-provisioning on higher-end devices.
    
    \item Scarce resources necessitate decentralized SLOiDs evaluation. The limited processing power, memory, and network bandwidth of CC devices make centralized, real-time monitoring, and evaluation of SLOiDs across the entire CC impractical or impossible. Constantly streaming detailed metrics from numerous constrained devices would overwhelm both the devices themselves (consuming scarce resources) and the network (causing congestion and increasing latency). Consequently, SLOiDs evaluation must be performed locally on the device or at a nearby edge node. This hardware-imposed limitation forces SLOiDs definitions to rely on metrics that can be efficiently computed and evaluated locally, potentially sacrificing global consistency or fine-grained observability for the sake of feasibility. 
    
    \item Constrained hardware necessitates incorporating SLOiDs elasticity trade-offs. On many CC devices, particularly at the Edge and IoT layers, traditional cloud elasticity mechanisms like rapid scaling up/down or scaling out instances are often not viable due to strict hardware resource limits, power constraints, or physical deployment realities. To ensure service objectives, elasticity must involve adjustments in quality (e.g., reducing data fidelity, frame rate) or performance (e.g., accepting higher latency). Therefore, SLOiDs and elasticity strategies for services deployed on such hardware must be defined to anticipate and manage these trade-offs. This could involve allowing graceful degradation under certain conditions, explicitly acknowledging that the limitations of the hosting device may prevent consistently meeting a single, rigid target.
\end{itemize}

In summary, the heterogeneity and resource constraints of CC hardware are not just complicating factors; they are fundamental characteristics that dictate:
\begin{enumerate}
    \item What types of SLOiDs are best suited for each part of the continuum.
    \item How SLOiDs must be monitored and evaluated.
    \item How SLOiDs must be defined to incorporate necessary trade-offs.
\end{enumerate}

%
%
%

In the context of this work, fulfilling SLOiDs maintains the system in homeostasis, or \textit{equilibrium}~\cite{sedlak_equilibrium_2024}, as we called it in previous works. This is a preferred steady state that spans across numerous devices and applications; in particular, SLOiDs can create hierarchies where the fulfillment is dependent on multiple lower-level SLOiDs or processes~\cite{casamayor_pujol_deepslos_2024}.

\vspace{5pt}
\textbf{System Equilibrium:} A computing system is in equilibrium if and only if all its SLOiDs are fulfilled.
\vspace{5pt}

Otherwise, if any SLOiD is violated, the system has lost its equilibrium; through controlled usage of elasticity strategies, it is possible to return the system to its equilibrium state. For instance, imagine a service under high load that violates its latency SLOiD, an elasticity strategy might throttle incoming requests or deploy an additional instance to reduce load, thereby returning the service to its equilibrium state. Failing to intervene risks that the service breaks beyond any possibility to returning to an equilibrium state. Furthermore, the concept of equilibrium can be complex. A system might possess multiple stable equilibrium states, each with different performance or efficiency characteristics. Elasticity strategies could potentially be used not just for recovery, but also to explore and transition between these states to optimize system operation.

\subsection{Partially Observable Markov Decision Processes -- POMDPs}
\label{subsec:pomdps}

A Markov-decision-process (MDP) is a time-discrete control process in which actions lead to partially random outcomes due to the stochastic nature of the system. In general, it models the transition probability of a new state given the current state of the system and the action of the agent. POMDPs generalize this to systems where the agent cannot observe all states of the system directly, but only the generated outcomes, or observations. As a simple example, consider an action that reduces the number of CPU cores available to a process; while this will affect the performance of the process, the extent of the impact is uncertain, as it may depend on other variables that are not directly observable. To address this kind of uncertainty,  in our AIF agents for CC systems we will always specify the POMDP governing the service. 

%
 MDPs are time-discrete control processes in which actions lead to partially random outcomes due to the stochastic nature of the system. 

In general, MDPs model the transition probability to a new state given the system's current state and an agent's action; hence, MDPs satisfy the Markov property, as counts for POMDPs. 

In the context of AIF, a POMDP is defined by a 7-tuple (S, U, \textbf{B}, O, \textbf{A}, C, D), where:

\begin{itemize}
    \item $S$ is the set of states for the agent
    \item U is a discrete set of actions
    \item \textbf{B} are the state transition model
    \item O is the set of observations
    \item \textbf{A} are the conditional observation probabilities
    \item C is the set of preferred outcomes
    \item D is the belief over the initial agent's state.
\end{itemize}

For many challenging problems, representing the state $S$ as a single, atomic label is inadequate. Real-world situations often involve multiple interacting factors that influence the system’s evolution. For instance, whether a computation finishes on time might depend on both its computational complexity and the current availability of system resources. To handle such intricacies, we often employ a factored representation. In this approach, the overall state space $S$ is defined as the Cartesian product of multiple state factors or modalities: $S=S_1 \times S_2 \times \dots S_N$. Thus, a specific state $s$ is defined as tuple $s = (s_1, s_2, \dots, s_N)$, where each element $s_i$ represents the value of the i-th factor (e.g., resource level, computation complexity) within its respective domain $S_i$. This factored structure naturally leads to representations of the transition (\textbf{B}) and observation (\textbf{A}) matrices that capture the interplay between these factors, often exploiting conditional independencies for a more compact and interpretable model.

\paragraph{The state transition models (\textbf{B})} They define the dynamics of the system, modelling $p(s_{t+1} | s_t, u_t)$, which is the probability of transitioning to state $s_{t+1}$ given the current state $s_t$ and action $u_t$. In the simple case with a single state modality ($N=1$), \textbf{B} can be viewed as a 3-tensor of shape $(|S|, |S|, |U|)$, where $|S|$ is the number of states and $|U|$ is the number of actions.
When the state is factored into $N$ modalities, $s_t = (s_{t,1}, \dots, s_{t,N})$, representing the full transition function $p(s_{t+1} | s_t, u_t)$ as a single high-dimensional tensor becomes impractical both computationally and for descriptive purposes. The size of such a tensor would grow exponentially with the number of modalities $N$.
To manage this complexity, we utilize the factored nature of the state space to represent the transition dynamics \textbf{B} in a structured and compact manner \cite{boutilier_decision-theoretic_1999}. This is commonly achieved using a \textit{factored representation}, often based on a Dynamic Bayesian Network (DBN). The underlying assumption is that the state of a single modality at the next time step, $s_{t+1, i}$, typically depends only on a limited subset of state modalities at the current time step $t$ (denoted as its parents, $Pa(s_i) \subseteq \{s_{t,1}, \dots, s_{t,N}\}$) and the action $u_t$.
Therefore, instead of defining one large tensor \textbf{B}, the transition model is specified by defining, for each modality $i$, its Conditional Probability Table (CPT): $P(s_{t+1, i} | Pa(s_i), u_t)$. These CPTs describe the local dependencies and how each modality evolves based on its specific influencing factors. These dependencies can be informed by an analysis of the system's behavior, such as the interdependencies illustrated in Figure~\ref{fig:producer_schema}.
A precise definition of these CPTs requires identifying the relevant parent variables $Pa(s_i)$ for each $s_i$ and quantifying their influence. Fortunately, the AIF framework allows for learning these parameters from interactions. In this work, we explore both expert-based specification and AIF-learned approaches for constructing the parameters of the factored transition model (\textbf{B}), assessing their respective challenges, benefits, and limitations.
Notice that in Section~\ref{subsec:pomdps_def}, where the POMDPs for all agents are defined, we will specify the DBNs for each factor and the specific rules detailing the CPTs can be found at the Appendix. 

\paragraph{The conditional observation probabilities (\textbf{A})} They define the likelihood of receiving a specific observation $o_t$ given the system is in state $s_t$. That is, they model the distribution $p(o_t | s_t)$. This likelihood function is crucial for the agent, as its goal is typically to invert this mapping via inference (e.g., using Bayes' theorem) to estimate the hidden state $s_t$ based on the observed outcome $o_t$. In its matrix form, \textbf{A} typically has dimensions $(|O|, |S|)$, where $|O|$ is the number of possible observations and $|S|$ is the number of states.
In this work, however, we make a simplifying assumption of \textit{perfect sensing}, meaning the observations correspond directly to the underlying states ($o_t = s_t$). We justify this assumption based on two primary reasons pertinent to our application domain. First, for the system under study, we possess \textit{full observability} over the variables constituting the state representation. For example, we do not need to infer internal states like GPU usage from indirect measurements; we can directly monitor the relevant system parameters. Second, the precision of sensor data obtainable from modern computing devices is generally high relative to the granularity of our state representation. Since we employ categorical variables with reasonably large bins for discretization, the inherent measurement precision typically does not necessitate a complex sensor noise model within our POMDP framework.
Consequently, for the agents defined in Section~\ref{subsec:pomdps_def}, the observation likelihood \textbf{A} reduces to an identity mapping. In the context of our factored state space $s = (s_1, \dots, s_N)$, this means that for each modality $i$, the observation $o_i$ perfectly reveals the state $s_i$, making the observation function an identity matrix for each factor.
It is important to acknowledge that this perfect sensing assumption should be relaxed when modeling more complex scenarios involving significant partial observability or latent variables that must be inferred indirectly. Examples include tasks requiring the inference of hidden user intentions or internal states of external systems (e.g., inferring the success of a remote machine learning service based on subsequent user interactions). For such problems, defining a non-identity \textbf{A} matrix becomes essential.

\paragraph{Preferred Outcomes (C)} This refers to a subset of possible outcomes that the agent aims to observe as a result of its actions. In our case, the fulfillment of SLOiDs is clearly a preferred observation for the agents. Note that preferred outcomes are not represented as probability distributions, but rather as log-probabilities. The preferred outcomes can be adjusted as the agent and the environment evolve, showing that preferences can change according to the current states.

\paragraph{Initial state (D)} This component encodes the agent’s belief over its initial state for each state modality. In this work, the initial state is randomized, but it can be specified to enforce particular scenarios if needed.

\subsection{Active Inference -- AIF}
AIF is a corollary of the Free Energy Principle (FEP), that describes how agents resolve uncertainty in their understanding of the world by minimizing their variational free energy~\cite{parr_active_2022}.
More generally, the FEP applies to all kinds of adaptive agents that learn accurate world models, also called \textit{generative models}, because they allow agents to understand and interpret observations \textit{generated} by unobservable latent variables. Hence, agents are able to minimize the discrepancies between their generative model and the actual generative process that governs the environment~\cite{friston_life_2013}. To minimize their free energy, agents take epistemic actions that allow them to resolve uncertainty, but they also change the environment to move it into a state that fulfills its internal preferences, achieving homeostasis. This describes the exploration-exploitation tradeoff inherent to agent-based systems. For more details about the FEP, the interested reader can check the following references \cite{friston_active_2016,kirchhoff_markov_2018,friston_reinforcement_2009,friston_designing_2024}.

In AIF, agents compute the expected free energy (EFE) for different policies ($\pi$), which are sequences of actions considered over various planning horizons. This allows agents to simulate their plan by computing the EFE for each action in the policy and selecting the one that minimizes EFE over the entire policy. In that regard, the policy length (\textit{pl}) represents the number of time steps ahead that the agent simulates when evaluating policies. A longer \textit{pl} allows the agent to consider more downstream consequences of actions when minimizing the EFE over the entire policy horizon. This means the optimal policy selected might involve an initial action that is not myopically optimal but enables lower EFE overall. However, increasing \textit{pl} significantly increases the computational resources required for planning. The EFE can be broken down into two main components:
\begin{equation}
\label{eq:efe}
\text{EFE} = - \overbrace{\mathbb{E}_{Q(o|\pi)} [\ln P(o|C)]}^{\text{Pragmatic Value}} -  \overbrace{\mathbb{E}_{Q(o,s|\pi)}D_{\mathrm{KL}} \left[ Q(s|o) \, \| \, Q(s) \right]}^{\text{Information Gain}}
\vspace{3pt}
\end{equation}
Here, the pragmatic value (\textit{pv}) reflects the expected log probability, given the approximated world model (or, the posterior distribution) $Q$, of obtaining a desirable observation $o$ under a specific policy $\pi$. The information gain (\textit{ig}) represents the expected reduction in uncertainty (that is, the expected model improvement), expressed as the expected Kullback-Leibler divergence ($D_{KL}$) between the prior belief about the world state $s$, and the posterior belief computed after obtaining a new observation. By balancing these factors, a policy that minimizes EFE not only achieves desired outcomes in the short term, but also enhance the agent's understanding of the environment, leading to improved decision-making in the longer term. Hence, AIF ensures an accurate model for decision-making, allowing agents to persist over time~\cite{palacios_markov_2020}. Specifically for CC systems, this ensures long-term SLOiD fulfillment. This capability, closely related to \textit{lifelong learning}, allows AIF agents to make decisions under uncertainty or dynamically changing environments. In particular, continuous exploration would foster alternative ways of scaling a service, and thus, increases its resilience.

\section{Vision}
\label{sec:vision}
Large scale computing systems, such as the CC, require mechanisms that underwrite each component's requirements in a decentralized manner. In this context, we want to highlight the term \textit{distributed intelligence}, as it encompasses numerous concepts that are essential for achieving an equilibrium in distributed computing systems. 

\subsection{Distributed Intelligence}

\begin{figure}[ht!]
    \centering
    \includegraphics[width=0.9\linewidth]{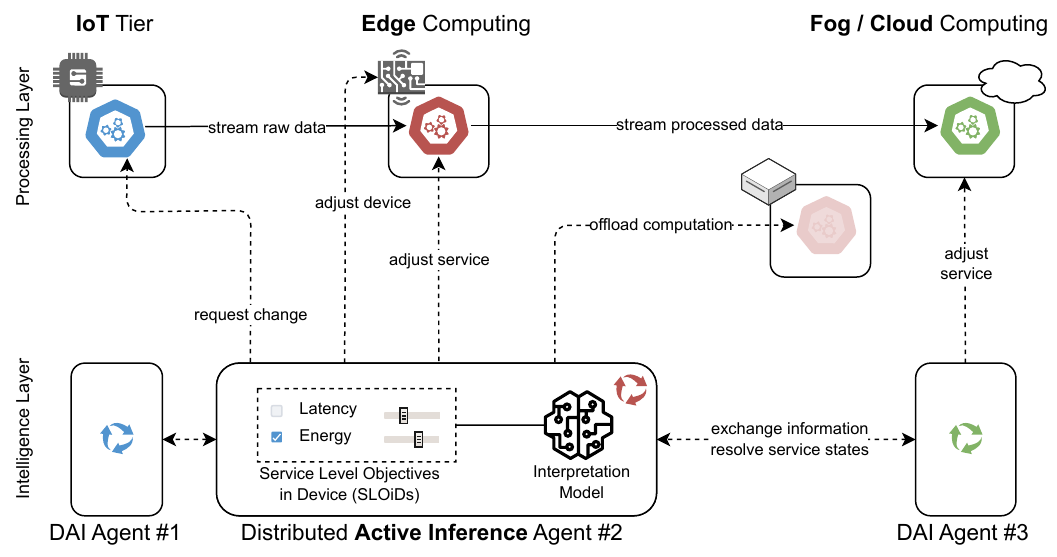}
    \vspace{6pt}
    \caption{High-level vision of distributed intelligence in computing continuum systems; each processing service is supervised by a distributed AIF agent that evaluates on-device requirements, communicates with other AIF agents to exchange information and resolve states of other components, and takes action whenever SLOiDs are violated to restore the system equilibrium.}
    \label{fig:vision}
\end{figure}

Contrarily to central Cloud services, the logic in CC systems is distributed over a widespread computing infrastructure. Figure~\ref{fig:vision} gives an intuition of a CC architecture that comprises multiple tiers, such as the IoT, and multiple computing layers from Edge, over Fog, up to the Cloud. As data travels from the IoT towards larger data centers, devices along the streaming pipeline can process the data through a network of microservices. Each processing service is supervised by a dedicated AIF agent that is responsible for (1) continuously evaluating SLOiD fulfillment and (2) taking actions whenever SLOiDs are violated, e.g., scaling the resources or quality of services. For taking actions, each agent uses a generative model that allows it to interpret the processing environment. Further, AIF agents (3) communicate and collaborate among themselves to exchange information (e.g., learned models or context) or actual workloads.

\subsection{Problem description}
\label{subsec:usecase}
Consider an application deployed within the CC that consists of three services organized in a pipeline. Each service performs a specific task, but they are interdependent, forming a sequential workflow. Using AIF agents, we will show how these services can autonomously adapt in a decentralized manner while cooperating to help each other achieve their objectives.

\begin{figure}[ht!]
    \centering
    \includegraphics[width=0.75\linewidth]{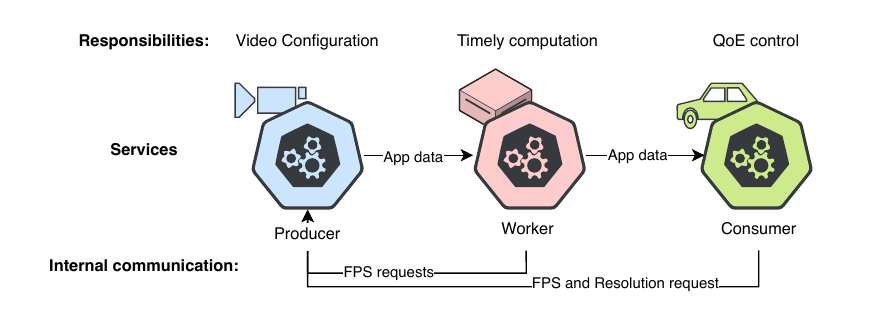}
    \caption{Smart city service pipeline from data producer to consumer. The \p provides a video stream and adapts its configuration according to the needs of the \w or \co. The \w performs latency-sensitive processing, while the \co ensures QoE by influencing the configuration of upstream components.}
    \label{fig:services_pipeline}
\end{figure}

The use case consists of three services, which we refer to as the \p, \w, and \co, according to their position in the processing pipeline. Consider that the three services can form a chain in a smart city use case, where each service is hosted on different devices distributed across a city; this is also depicted in Figure~\ref{fig:services_pipeline}:

The \p service is attached to a video camera, interacting with the camera to prepare and send batches of images to a downstream service for further processing. 
The \w service, deployed on an Edge server, performs tasks such as face detection and blurring to preserve pedestrian anonymity. Once the images are processed, they are sent to the final user of the image stream.
The \co service runs on a smartphone or in an autonomous vehicle for traffic analysis, ensuring that the client’s quality of experience (QoE) requirements are met to maintain user satisfaction with the overall application.

Each service must fulfill a different set of SLOiDs and have different actions available.
The \co SLOiDs are focused on controlling the user’s Quality of Experience (QoE). 
Three are defined: the first, named \slos, ensures that the service has correctly fulfilled its goal, e.g., the correct application of a privacy filter to each video frame. The second, \slod, measures the quality of the video stream by assessing the smoothness of consecutive frames. 
The third SLOiD, \slocc, monitors the energy consumption of the service, including the extra cost needed to communicate with other services.
The \co service has a single action available (\textit{Toggle\_comm}), which consists of enabling or disabling the communication channel with the \p service in order to request a change on the video stream configuration, i.e., changing the frames per second or the resolution.

The \w processes the video stream as a batch of images received within a specified deadline. The SLOiD \sloi, ensures that the processing time for each batch remains below a given threshold. Like the \co, the \w also has a \slowc SLOiD, which tracks its energy consumption and the overhead associated with communicating with other services. Note that in both cases, communication refers to non-functional data exchange that improves control over SLOiDs.
The \w has two actions available: it can switch on and off its GPU in order to accelerate the images processing (\textit{Switch\_GPU}), notice that this is only possible when the deployment host has an available GPU. Second, it can enable or disable the communication channel (\textit{Toggle\_comm}) with the \p service to request a change on FPS configuration of the video stream.

The \p records the video stream with specific configurations and sends the images to the \w. As the service with direct control over the recording settings, the \p’s non-functional objective is to adjust the video configuration to satisfy both the \w and the \co. Therefore, its three SLOiDs are \textbf{Worker satisfaction with FPS} (\slowd), \textbf{Consumer satisfaction with FPS} (\slocdf), and \textbf{Consumer satisfaction with Resolution} (\textbf{\slocdr}). When the \p does not receive any request from the \w or the \co or the request is not to change anything, the SLOiDs are considered fulfilled. 
The \p has 2 possible actions which consist of changing the resolution (\textit{Change\_resolution}) and the FPS (\textit{Change\_FPS}) of its attached camera. 

\begin{table}[h!]
\centering
\caption{Summary of the defined SLOiDs for the three AIF agents}
\label{tab:sloids_summary}
\resizebox{\textwidth}{!}{%
\begin{tabular}{lclclc}
\multicolumn{2}{c}{\p} & \multicolumn{2}{c}{\w} & \multicolumn{2}{c}{\co} \\
\multicolumn{1}{|l}{\textit{Description}} & \multicolumn{1}{c|}{\textit{}} & \textit{Description} & \multicolumn{1}{c|}{\textit{}} & \textit{Description} & \multicolumn{1}{c|}{\textit{}} \\
\multicolumn{1}{|l}{Worker satisfaction with FPS} & \multicolumn{1}{c|}{\slowd} & Processing time & \multicolumn{1}{c|}{\sloi} & Correct application of the privacy filter & \multicolumn{1}{c|}{\slos} \\
\multicolumn{1}{|l}{Consumer satisfaction with FPS} & \multicolumn{1}{c|}{\slocdf} & Energy consumption of the service & \multicolumn{1}{c|}{\slowc} & Quality of the final video stream & \multicolumn{1}{c|}{\slod} \\
\multicolumn{1}{|l}{Consumer satisfaction with resolution} & \multicolumn{1}{c|}{\slocdr} &  & \multicolumn{1}{c|}{} & Energy consumption of the service & \multicolumn{1}{c|}{\slocc}
\end{tabular}%
}
\end{table}

\section{Methodology}
\label{sec:methodology}
In this article, we demonstrate how intelligence can be distributed across CC services by modeling agents using AIF, enabling them to manage services autonomously and collaboratively, an essential capability for decentralized application management in the CC.
Now, we formally define the AIF agents as POMDPs, describe the tools and simulation setup, and conclude with a description of the experiments.

\subsection{POMDPs definition}
\label{subsec:pomdps_def}

\subsubsection{\textbf{\p's POMDP model}}

\begin{figure}[ht!]
    \centering
    \includegraphics[width=0.9\linewidth]{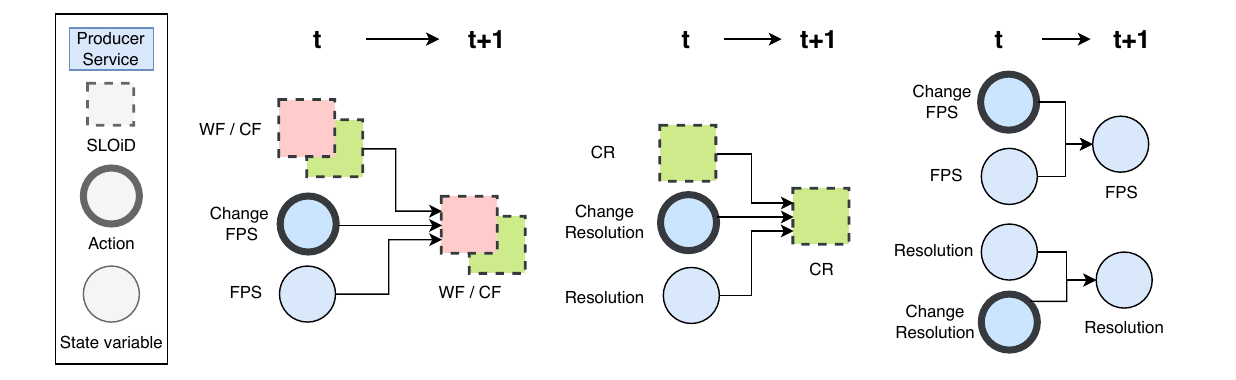}
    \caption{\p’s DBN. The model includes five modalities, two of which behave identically and are therefore grouped together in the figure. The colors correspond to the services each modality is associated with.}
    \label{fig:producer_schema}
\end{figure}

\paragraph{State modalities (S)}
The \p's state ($S$) contains five modalities. Three of them are SLOiDs: \slowd, \slocdf, and \slocdr, as in Table \ref{tab:sloids_summary}. The other two state modalities are \textit{FPS} and \textit{Resolution}. The \slowd, \slocdf, and \slocdr modalities can each take three state values: \actInc, \actSt, and \actDec. The \actSt value is assumed when no message is received from either the \co or the \w, as if they are satisfied with the current value.  
The \textit{FPS} modality can take the values $\{12, 16, 20, 26, 30\}$, and the \textit{Resolution} modality can take the values $\{120p, 180p, 240p, 360p, 480p, 720p\}$. Figure \ref{fig:producer_schema} illustrates the modalities and actions of the \p, and how they influence the state modalities at the next time step. This can be interpreted as a Dynamic Bayesian Network (DBN) representing the evolution of the \p’s state modalities, with the assumption that the relationships between variables remain fixed at each time step.

\paragraph{Actions (U)}
The \p’s action set (U) consists of two actions, each with three possible values.
\begin{enumerate}
    \item \textit{Change\_FPS}: \actInc, \actSt, and \actDec the \textit{FPS} current value.
    \item \textit{Change\_resolution}: \actInc, \actSt, and \actDec the \textit{Resolution} current value.
\end{enumerate}

Consider that to limit the exponential growth of possible actions, both \actInc and \actDec modify the current value by one step, e.g., increasing resolution from $120p$ to $180p$.

\paragraph{State Transition Model (\textbf{B})}
The dependencies defining the transition probabilities \textbf{B} for each state modality, given other system states at time $t$ and actions $u_t$, are illustrated in Figure~\ref{fig:producer_schema}. We define the Conditional Probability Tables (CPTs) for each modality using the rules described in Appendix \ref{sec:appendix-producer}.
For the alternative case where the AIF agent learns the transition parameters, the CPTs are initialized with uniform probability distributions instead of the deterministic rules. However, the underlying DBN structure defining the dependencies (as shown in Figure~\ref{fig:producer_schema}) remains identical in both the expert-defined and learned models. This applies for the 3 POMDP models described, i.e., the \p, \w, and \co.

\paragraph{Observations (O)}
The set of observations represents the possible outputs of the external environment, produced after an action $u \in U$ has been taken. As explained earlier, in this work we are assuming a perfect sensing. That is, an identity mapping between states and observations. To this end, the set of observations (O) for the agent is equivalent to the set of states (S). Consequently the likelihood mapping $A$ is defined as a set of identity matrices, one per modality. This applies for the 3 POMDP models described.

\paragraph{Preferred Ouctomes (C)}
In this work, we consider the \p to have no preferences over the \textit{FPS} and \textit{Resolution} sets. The preferred outcomes of the requests from the \w and the \co are to receive a \actSt, meaning their configuration requests are satisfied. More in detail, the vectors are initially defined as follows: $C^{\slowd} = \{ 0.25, 1.5, 0.25\}; \; C^{\slocdf} = \{ 0.5, 3.0, 0.5\}; \; \text{and} \; C^{\slocdr} = \{ 0.5, 3.0, 0.5\}$. 
It is worth mentioning that the requests from the \co have a higher maximum value than the ones of the \w, to indicate the ultimate goal is achieving satisfaction at the end of the pipeline, and considering that the \w might have other capabilities to satisfy its goals. 

\paragraph{Initial state (D)}
We consider the initial state to be unknown, its values are uniformly distributed adding up to 1. This applies for the 3 POMDP models described.

\subsubsection{\textbf{\w's POMDP model}}
\begin{figure}[ht!]
    \centering
    \includegraphics[width=0.9\linewidth]{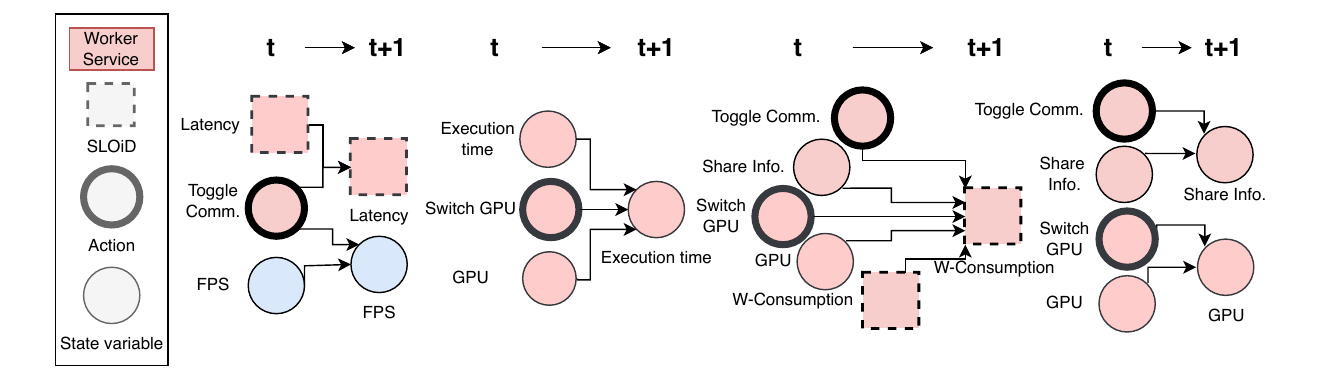}
    \caption{\w’s DBN. The model includes six modalities, two of which are SLOiDs. The colors correspond to the services each modality is associated with.}
    \label{fig:worker_schema}
\end{figure}

\paragraph{State modalities (S)}
The \w has six state ($S$) modalities. Two of them are SLOiDs: \sloi and \slowc, as in Table \ref{tab:sloids_summary}. The others are state variables: \textit{Execution Time}, \textit{FPS}, \textit{Share Information}, and \textit{GPU}, all can be seen in Figure \ref{fig:worker_schema}. Each modality can take specific values based on the system's behavior.  
\sloi is a boolean modality: it takes the value \textit{True} when the computation is completed within the deadline, and \textit{False} when the computation exceeds the deadline.  
\textit{Execution Time} is a categorical modality representing the time required to complete the computational task. It can take the values \{\textit{LOW, MID-LOW, MID, MID-HIGH, HIGH}\}, corresponding to the following durations in milliseconds: $(0\,,\,15], (15\,,\,30], (30\,,\,45], (45\,,\,60], \text{ and } (60\,,\,\infty)$.  
The \textit{FPS} state modality is defined as for the \p.  
\slowc is a state modality that reflects the energy consumption of the device, considering GPU usage and the activation of the upstream communication channel. The levels are \textit{LOW}, \textit{MID}, and \textit{HIGH}, corresponding to an overall consumption below 7W, between 7–8W, and above 8W, respectively.  
\textit{Share Information} is a boolean state modality that is \textit{True} when the upstream communication channel is active, allowing the \w to communicate with the \p, and \textit{False} when it is disabled.  
\textit{GPU} is also a boolean state modality: it is \textit{True} when the GPU is on and \textit{False} when it is off.

\paragraph{Actions (U)}
The \w's action set (U) consist of two actions:
\begin{enumerate}
    \item \textit{Switch\_GPU}: \actOff and \actOn modifies the GPU current state and \actSt keeps the current state.
    \item \textit{Toggle\_comm}: \actEn and \actDis the upstream communication channel. 
\end{enumerate}

\paragraph{State Transition Model (\textbf{B})}
The dependencies defining the transition probabilities \textbf{B} for each state modality, given other system states at time $t$ and actions $u_t$, are illustrated in Figure~\ref{fig:worker_schema}. We define the Conditional Probability Tables (CPTs) for each modality using the rules described in Appendix~\ref{sec:appendix-worker}.

\paragraph{Preferred Outcomes (C)}
There are no preferred outcomes for the \textit{FPS}, the \textit{Share Information}, or the \textit{GPU}, hence they are initialized as a vector of zeros. The \sloi preferred outcome is \textit{True}, hence the vector is defined $C^{\,\text{\sloi}} = \{ 0.1, 3\}$. The \textit{Execution Time} prefers values that are \textit{MID} or below, the vector is: $C^{\,\text{Exec. Time}} = \{3, 2.5, 2, 0.25, 0.1\}$. Finally, the \slowc is expected to be \textit{LOW} or \textit{MID}, its vector is: $C^{\,\text{\slowc}} = \{ 3, 2.5, 0,5 \}$


\subsubsection{\textbf{\co's POMDP model}}

\begin{figure}[ht!]
    \centering
    \includegraphics[width=0.9\linewidth]{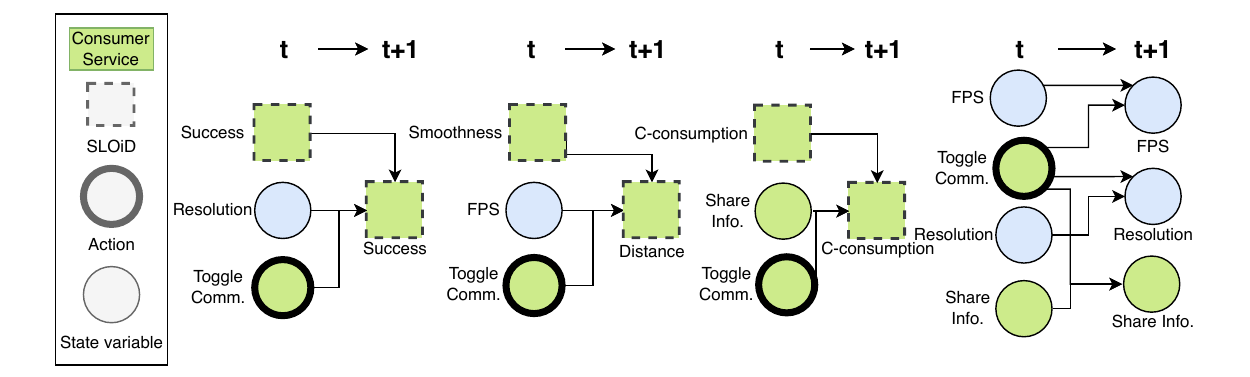}
    \caption{\co’s DBN. The model includes six modalities, three of which are SLOiDs. The colors correspond to the services each modality is associated with.}
    \label{fig:consumer_schema}
\end{figure}

\paragraph{State modalities (S)}
The \co's state ($S$) consists of six modalities. Three of them are SLOiDs: \slos, \slod, and \slocc, as in Table \ref{tab:sloids_summary}. The other three are system variables: \textit{FPS}, \textit{Resolution}, and \textit{Share Information}, as can be seen in Figure \ref{fig:consumer_schema}. Each modality can take specific values based on the system's behavior.  
\slos is a boolean modality: it takes the value \textit{True} if the service has properly applied the privacy filter, and \textit{False} otherwise.  
\slod is a categorical state modality representing the distance shift of a pixel between two consecutive frames. It can take the values (\textit{SHORT, MID-SHORT, MID, MID-LONG, LONG}), which correspond to a pixel movement of $(0\!-\!25], (25, 50], (50,75], (75,100], (100, \infty)$ pixels.  
The \slocc modality is defined as for the \w, but without the influence of the GPU status, only considering the device's energy consumption and the activation of upstream communication.
Both the \textit{FPS} and the \textit{Resolution} state modalities are defined similarly to those for the\p. 
Finally, the \textit{Share Information} is defined in the same way as for the \w.

\paragraph{Actions (U)}
The \co action set (U) consist of:
\begin{enumerate}
    \item \textit{Toggle\_comm}: \actEn and \actDis the upstream communication channel. 
\end{enumerate}

\paragraph{State Transition Model (\textbf{B})}
The dependencies defining the transition model \textbf{B} for each state modality, given other system states at time $t$ and actions $u_t$, are illustrated in Figure~\ref{fig:consumer_schema}. We define the Conditional Probability Tables (CPTs) for each modality using the rules described in Appendix~\ref{sec:appendix-consumer}.

\paragraph{Preferred Outcomes (C)}
There are no preferred outcomes for the \textit{FPS}, \textit{Resolution}, or \textit{Share Information}, so their vectors are all zeros. The \slos preferred outcome is \textit{True}, hence the vector is as follows: $C^{\,\text{\slos}}=\{0.25, 3\}$. The \slod preferred outcome is MID or lower, so the vector is: $C^{\,\text{\slod}}=\{3, 2.5, 2, 0.5, 0.1\}$. Finally, the \slocc preferred outcome is MID or lower, so the vector is: $C^{\,\text{\slocc}}=\{3, 2.5,0.5\}$.

\subsection{Tools and simulation setup}
In this section we present the AIF library used, the dataset, the key elements of the simulation process.

\subsubsection{Active inference with \textit{pymdp}}
\label{subsec:pymdp}
All experiments in this article are conducted using \textit{pymdp}~\cite{heins_pymdp_2022}, which is a Python library that simplifies the implementation of AIF agents. \textit{pymdp} provides a set of classes and methods for modeling, inference, learning, and control: the core class, \texttt{Agent}, allows to define an AIF agent, whose methods provide the key functionalities needed for AIF agents, such as policy/action selection, state inference, and parameter learning. 

It is important to note that the type of actions performed by the agents, as well as the complex relationships between state modalities and actions, are not commonly encountered in standard AIF scenarios. Therefore, the version of \textit{pymdp} used for this work was based on a customized fork from GitHub\footnote{https://github.com/ran-wei-verses/pymdp}, which is specifically adapted to handle these types of complex state transitions.

\subsubsection{Dataset}
The dataset are the traces of the processing environment of the pipeline of streaming services. The traces have been recorded using different hardware to account for the diversity of the CC.
The dataset records various system parameters, including execution time, CPU utilization, memory usage, energy consumption, image resolution, frames per second (FPS), the success of the privacy model, the smoothness between consecutive frames, the type of device, and GPU usage in the described pipeline. So, it provides all the required data to rebuild the use case for this article. 
The dataset is publicly available on GitHub\footnote{\href{https://github.com/borissedlak/analysis/tree/main/data}{https://github.com/borissedlak/analysis/tree/main/data}, last accessed on April 25th, 2025}.

The use case described here is similar to that presented in \cite{sedlak_equilibrium_2024}, in the sense that it consists of multiple sequential processing services.
However, whereas in \cite{sedlak_equilibrium_2024} the AIF agent controls the entire system, in this paper, each service is associated with an autonomous agent, making the system both multi-agent and decentralized.

Further, the dataset was generated by independently adjusting key system parameters —image resolution, video frame rate, and GPU status— providing an independent set of behaviors for each parameter combination. This makes the dataset particularly well-suited for our simulation experiments, as it allows us to model independent behaviors and interactions between variables. 

The dataset is used to simulate different operational scenarios for the pipeline, helping to validate the performance of the active inference agents under varying conditions. Since the dataset missed few combinations of resolution and FPS, some missing FPS values were approximated. Specifically, the missing FPS values were interpolated by averaging the neighboring values. The standard deviation of these neighboring values was also considered to ensure a realistic approximation of the missing data. 

\subsubsection{Simulation}
Several simulation experiments are conducted to identify the gaps needed to leverage active inference in the context of Computing Continuum system, to make distributed intelligence a reality. 
Combining both disciplines (AIF and the CC) presented various challenges, and running experiments helped us to better understand how they can be integrated effectively. The code for these simulations can be found in the following GitHub Repo\footnote{https://github.com/vikcas/AIF\_for\_CC}.

Each experiment was run for 200 time steps and repeated at least 10 times to account for the variations in initial conditions. These correspond to the initial values of the system's parameters: image \textit{Resolution}, \textit{FPS}, \textit{GPU} status, \w sharing information, and \co sharing information. Randomized initial configurations are used in every repetition ensuring that agents could start in diverse states, including possible equilibrium states where no immediate action is required.

The simulation process can be summarized as follows:
\begin{enumerate}
    \item \textbf{Initialization.} At the beginning of each experiment, the agents' initial state and parameters are set randomly.
    \item \textbf{Policy computation.} Each agent computes the best policy according to the active inference framework, i.e., a policy that minimizes the EFE. 
    \item \textbf{Action selection.} The agents selects the first action and applies it to the system. Consequently, the environment and the system changes. It is important to remark that agents can take all available actions at each time steps, meaning that the \p can simultaneously change the resolution and the FPS of the camera.
    \item \textbf{New Observation.} The environment generates a new observation for each agent by sampling from the dataset; this sample need not follow a temporal sequence but must correspond to the result of the chosen action.
    \item \textbf{Infer state and learn parameters.} Based on these new observations, agents can infer their new state and learn the parameters of the transition model.
    \item \textbf{Repetition.} At this point a new policy is computed, repeating the cycle for 200 times.
\end{enumerate}


\subsection{Evaluation}
During the experiments, we evaluated two key aspects:
\begin{itemize}
    \item SLOiDs fulfillment rate. The agent's ability fo meet its service goals. For visualization purposes, at each time step the cumulative SLOiD fulfillment rate is shown.
    \item Expected Free Energy. A measure used by AIF to assess their performance in terms of selecting the action that helps them improve their model while fulfilling the SLOiDs, see Equation~\eqref{eq:efe}.
\end{itemize}

Each agent has the following SLOiDs fulfillment criteria:

\begin{itemize}
    \item \p: \slowd, \slocdf, and \slocdr = \{ \actSt\}
    \item \w: \slowc $\leq$ \{ \textit{MID} \} and \sloi = \{ \textit{TRUE} \}
    \item \co: \slos = \{ \textit{TRUE} \}, \slod and \slocc $\leq$ \{ \textit{MID} \}
\end{itemize}

\subsubsection{Experiments}
We conduct five experiments to assess the suitability of AIF agents for autonomous and distributed management of SLOiD fulfillment across the service pipeline.

\paragraph{SLOiD fulfillment} This experiment evaluates how the three AIF agents fulfill their respective SLOiDs by selecting appropriate actions given each state of the environment. We fix the policy length to 3 and use the defined transition model. Results are presented in Section~\ref{res:slo_fulfill_rate}.

\paragraph{Learning the transition model} This experiment assesses the impact of learning the transition model (\textbf{B}) on the agent's performance. We again set the policy length to 3, use the default learning parameters from the \textit{pymdp} library, and initialize the transition model with \textbf{B} matrices, each containing a uniform distribution over transitions from any current state to the next. Results are presented in Section~\ref{res:slo_fulfill_rate_learning}.


\paragraph{Comparison to multi-agent reinforcement learning} This experiment uses the same environment than the 2 previous experiments, but the SLOiDs are controlled by 3 reinforcement learning-based agents. We use an algorithm from the state-off-the-art named Proximal Policy Optimization \cite{schulman_proximal_2017} and compare the results with the previously obtained as a baseline for the AIF agents. Results are presented in Section~\ref{subsec:marl_comp}.



\paragraph{Heterogeneous hardware}
This experiment explores the response of the AIF agents to severe environment changes, mimicking as if they were redeployed in different hardware. To do so, after 75 time-steps into controlling the services in a specific device, they are change to a different one. We see both the dynamic adaptation of the system, as well as the need of adapting the SLOiDs to the pair service-hardware. Results are presented in Section~\ref{subsec:het_hw}

\paragraph{Computing cost}
This experiment analyzes the computational cost and the resulting performance of AIF agents. We compare the initial results obtained to those obtained with a shorter policy length. We have seen that given the same models, the policy length has a large effect on the computational cost. Results are presented in Section~\ref{subsec:comp_cost}.

\section{Results}
\label{sec:results}

\subsection{SLOiD fulfillment}
\label{res:slo_fulfill_rate}
\begin{figure}[ht!]
  \centering
\begin{subfigure}[t]{.315\linewidth}
    \centering\includegraphics[width=1.0\linewidth]{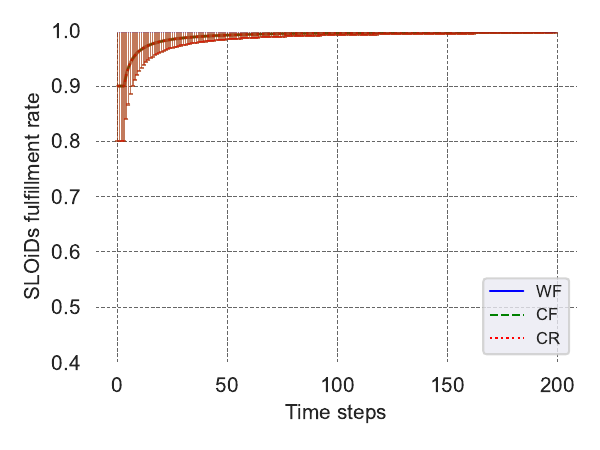}
    \captionsetup{skip=0pt}
    \caption{\p SLOiDs fulfillment rate for \slowd, \slocdf, and \slocdr.}
    \label{fig:nl_pl3_slo_p}
\end{subfigure}
\hspace{.015\linewidth}
\begin{subfigure}[t]{.315\linewidth}
    \centering\includegraphics[width=1.0\linewidth]{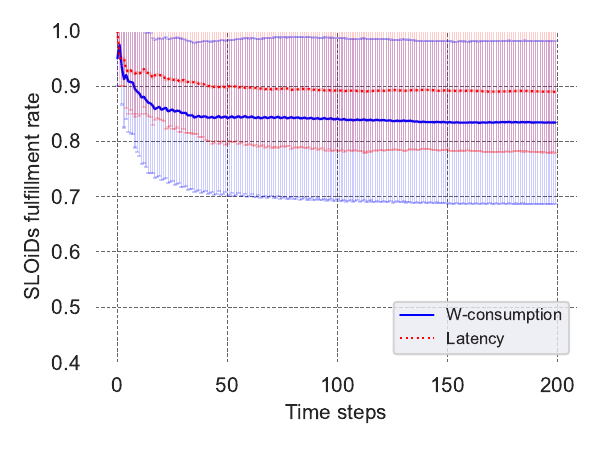}
    \captionsetup{skip=0pt}
     \caption{\w SLOiDs fulfillment rate for \sloi and \slowc.}
     \label{fig:nl_pl3_slo_w}
\end{subfigure}
\hspace{.015\linewidth}
\begin{subfigure}[t]{.315\linewidth}
    \centering\includegraphics[width=1.0\linewidth]{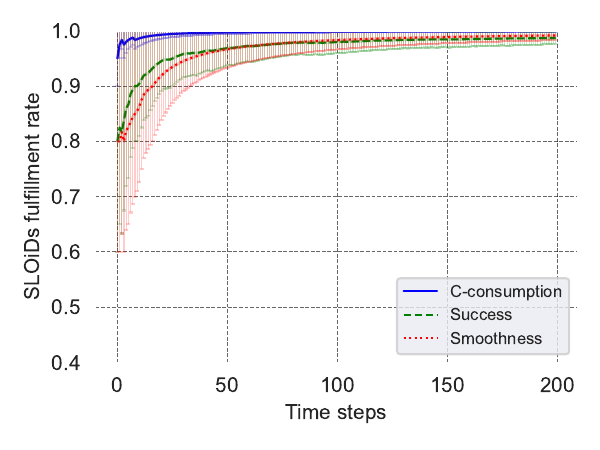}
    \captionsetup{skip=0pt}
    \caption{\co SLOiDs fulfillment for \slos, \slod, and \slocc.}
    \label{fig:nl_pl3_slo_co}
\end{subfigure}
\vspace{6pt}
\caption{SLOiDs fulfillment rate with no \textbf{B} parameter learning and policy length of 3.}
\vspace{-4pt}
\label{fig:nl_pl3_slos}
\end{figure}

This experiment assesses the agents' ability to select appropriate actions based on expert-defined transition models. As shown in Figure~\ref{fig:nl_pl3_slos}, all agents stabilize their SLOiDs fulfillment rates around 50 steps. However, the \w agent clearly struggles more to achieve high SLOiDs fulfillment rates compared to the others. This reduced performance is primarily due to the competing SLOiDs integrated into its model. Specifically, the \slowc mandates switching off the GPU and communication channel, while the \sloi requires them to be switched on if the proper setting of the camera is not selected. 
In addition to its lower maximum SLOiD fulfillment rate, the \w agent also exhibits greater performance variance. This higher variance may stem from the \p’s preference for prioritizing the satisfaction of the \co over that of the \w. As a result, even if the \w consistently selects optimal actions, the outcome may not meet expectations, depending on the requests from the \co to the \p.

\subsection{Learning the transition model}
\label{res:slo_fulfill_rate_learning} 
\begin{figure}[ht!]
  \centering
\begin{subfigure}[t]{.315\linewidth}
    \centering
    \includegraphics[width=1.0\linewidth]{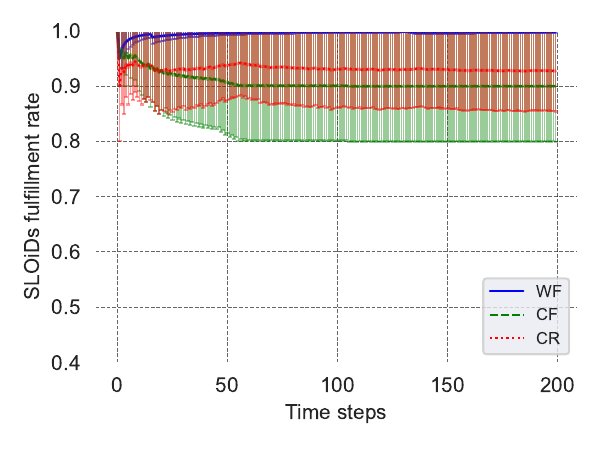}
    \captionsetup{skip=0pt}
    \caption{\p SLOiDs fulfillment rate for \slowd, \slocdf, and \slocdr.}
    \label{fig:l_pl3_slo_p}
\end{subfigure}
\hspace{.015\linewidth}
\begin{subfigure}[t]{.315\linewidth}
    \centering
    \includegraphics[width=1.0\linewidth]{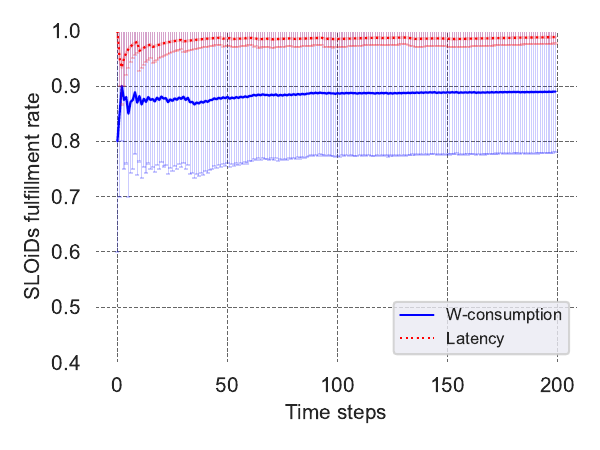}
    \captionsetup{skip=0pt}
     \caption{\w SLOiDs fulfillment rate for \sloi and \slowc.}
     \label{fig:l_pl3_slo_w}
\end{subfigure}
\hspace{.015\linewidth}
\begin{subfigure}[t]{.315\linewidth}
    \centering
    \includegraphics[width=1.0\linewidth]{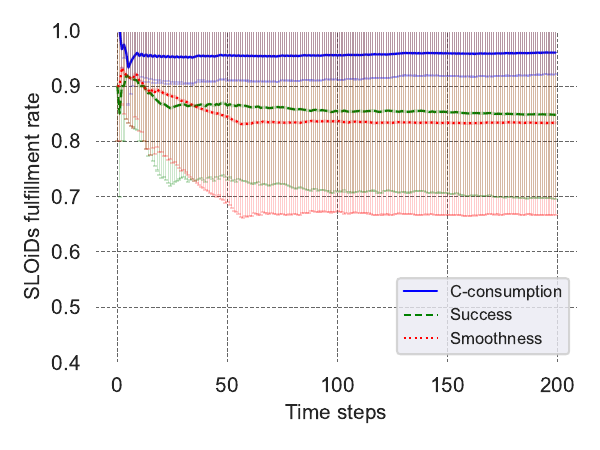}
    \captionsetup{skip=0pt}
    \caption{\co SLOiDs fulfillment for \slos, \slod, and \slocc.}
    \label{fig:l_pl3_slo_co}
\end{subfigure}
\vspace{6pt}
\caption{SLOiDs fulfillment rate with \textbf{B} parameter learning and policy length of 3.}
\vspace{-4pt}
\label{fig:l_pl3_slos}
\end{figure}

Figure~\ref{fig:l_pl3_slos} shows the SLOiD fulfillment rate for each AIF agent in a variation of the previous experiment where agents must select actions to fulfill SLOiDs while simultaneously learning the transition model. 
In comparison with the results from the previous experiment (Figure~\ref{fig:nl_pl3_slos}), we observe that all agents stabilize their SLOiD fulfillment rates also around the 50 time steps. The overall fulfillment rate of the agents is lower than in the previous case. This indicates increased difficulty in finding performance-maximizing policies when the model is unknown and must be learned online. For example, agents \p and \co, which reached nearly 100\% fulfillment in the previous experiment, are around 80-90\% in this learning scenario. An interesting observation is that in these learning trials, the \w agent appears to prioritize the \sloi over \slowc, resulting in better fulfillment rates compared to its performance in the previous experiment with the expert model.

\subsection{Comparison to multi-agent reinforcement learning (MARL)}
\label{subsec:marl_comp}
The same three agents were implemented within a multi-agent reinforcement learning (MARL) framework using RLlib \cite{liang_rllib_2018, wu_rllib_2021} and the PPO algorithm \cite{schulman_proximal_2017}. Consistent with the AIF experiments, we did not fine-tune any parameters, allowing us to assess each method's general capabilities in this scenario. We consider two distinct training regimes: minimal ($400{,}000$ time steps) and extensive ($40$ million time steps), mirroring the exploration of two scenarios with the AIF agents. Evaluation is performed over $200$ time steps.
The reward function of each agent is computed by assigning a reward for every SLOiD it fulfills. Therefore, actions that lead to fulfilling more SLOiDs result in higher rewards.

\begin{figure}[ht!]
  \centering
\begin{subfigure}[t]{.315\linewidth}
    \centering
    \includegraphics[width=1.0\linewidth]{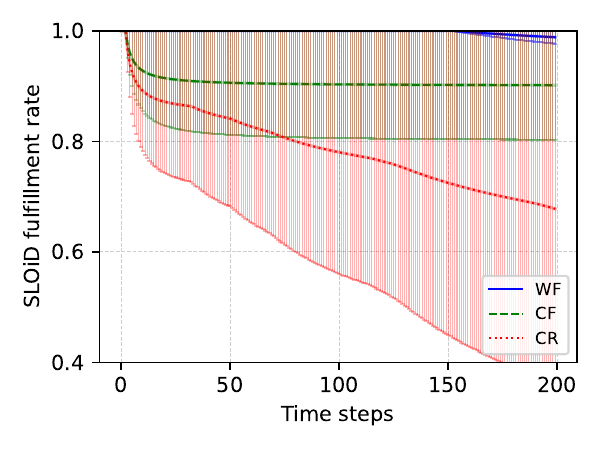}
    \captionsetup{skip=0pt}
    \caption{Minimally trained MARL \p agent. SLOiDs fulfillment rate for \slowd, \slocdf, and \slocdr.}
    \label{fig:marl_min_producer}
\end{subfigure}
\hspace{.015\linewidth}
\begin{subfigure}[t]{.315\linewidth}
    \centering
    \includegraphics[width=1.0\linewidth]{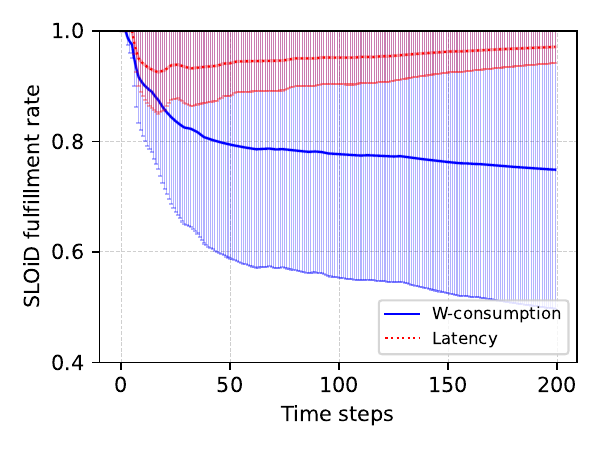}
    \captionsetup{skip=0pt}
     \caption{Minimally trained MARL \w agent. SLOiDs fulfillment rate for \sloi and \slowc.}
     \label{fig:marl_min_worker}
\end{subfigure}
\hspace{.015\linewidth}
\begin{subfigure}[t]{.315\linewidth}
    \centering
    \includegraphics[width=1.0\linewidth]{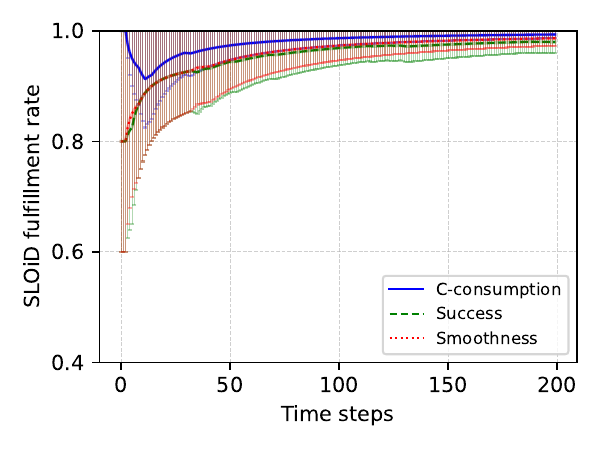}
    \captionsetup{skip=0pt}
    \caption{Minimally trained MARL \co agent. SLOiDs fulfillment for \slos, \slod, and \slocc.}
    \label{fig:marl_min_consumer}
\end{subfigure}
\begin{subfigure}[t]{.315\linewidth}
    \centering
    \includegraphics[width=1.0\linewidth]{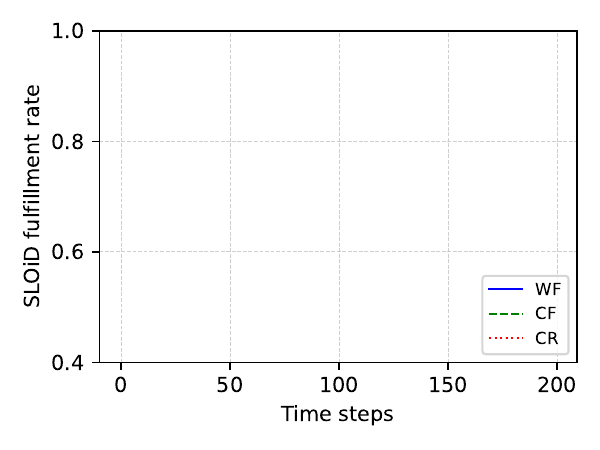}
    \captionsetup{skip=0pt}
    \caption{Extensively trained MARL \p agent. SLOiDs fulfillment rate for \slowd, \slocdf, and \slocdr.}
    \label{fig:marl_well_producer}
\end{subfigure}
\hspace{.015\linewidth}
\begin{subfigure}[t]{.315\linewidth}
    \centering
    \includegraphics[width=1.0\linewidth]{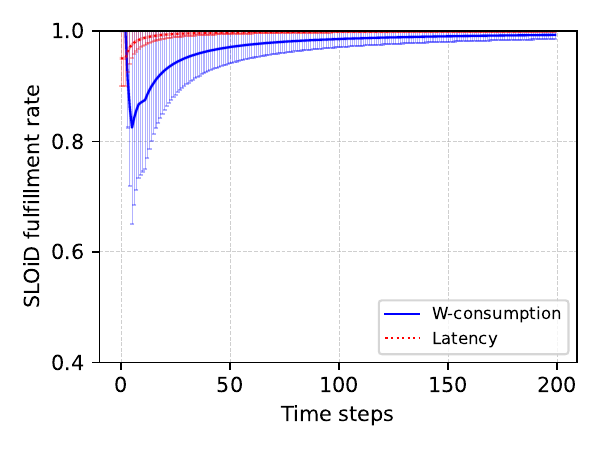}
    \captionsetup{skip=0pt}
     \caption{Extensively trained MARL \w agent. SLOiDs fulfillment rate for \sloi and \slowc.}
     \label{fig:marl_well_worker}
\end{subfigure}
\hspace{.015\linewidth}
\begin{subfigure}[t]{.315\linewidth}
    \centering
    \includegraphics[width=1.0\linewidth]{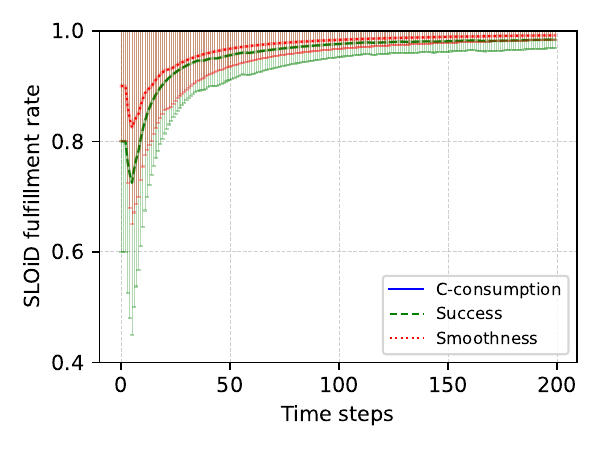}
    \captionsetup{skip=0pt}
    \caption{Extensively trained MARL \co agent. SLOiDs fulfillment for \slos, \slod, and \slocc.}
    \label{fig:marl_well_consumer}
\end{subfigure}
\vspace{6pt}
\caption{MARL simulation, showing the SLOiD fulfillment rates for the minimally- and well-trained agents}
\vspace{-4pt}
\label{fig:MARL}
\end{figure}

Figure~\ref{fig:MARL} shows the SLOiD fulfillment rates for all the MARL agents. The top three plots (Figures~\ref{fig:marl_min_producer}, \ref{fig:marl_min_worker}, and~\ref{fig:marl_min_consumer}) correspond to agents trained under the minimal training regime. In this case, their performance is clearly lower than that of the AIF agents. The \p agent fails to stabilize consumer satisfaction with resolution, and the \w agent shows a declining trend in fulfilling the \slowc SLOiD.
In contrast, the three plots at the bottom (Figures~\ref{fig:marl_well_producer}, \ref{fig:marl_well_worker}, and~\ref{fig:marl_well_consumer}) correspond to the extensive training regime, where agents have acquired sufficient knowledge of the environment to consistently fulfill all their SLOiDs.

Although this article does not aim for a direct comparison between the two approaches, it is worth noting the significant data efficiency of the AIF agents. Specifically, the learning AIF agents achieved their results with less than $200$ time steps of runtime data, a stark contrast to MARL agents which model the environment through neural networks, which typically require large amounts of data.

\subsection{Heterogeneous hardware}
\label{subsec:het_hw}
\begin{figure}[ht!]
  \centering
\begin{subfigure}[t]{.4\linewidth}
    \centering
    \includegraphics[width=0.75\linewidth]{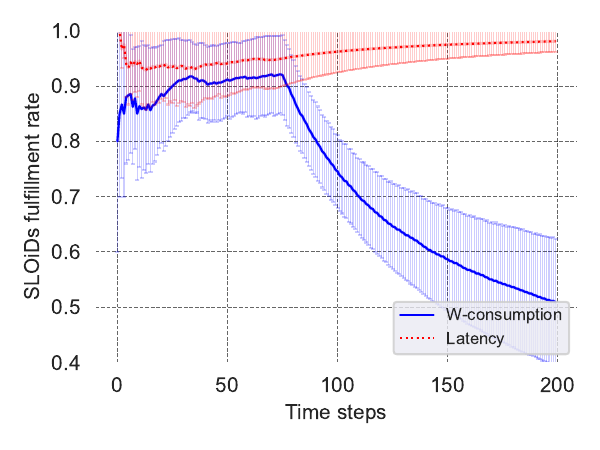}
    \captionsetup{skip=0pt}
    \caption{\w SLOiDs fulfillment while learning transition model and facing sudden environmental change.}
    \label{fig:env_change_l}
\end{subfigure}
\hspace{1cm}
\begin{subfigure}[t]{.4\linewidth}
    \centering
    \includegraphics[width=0.75\linewidth]{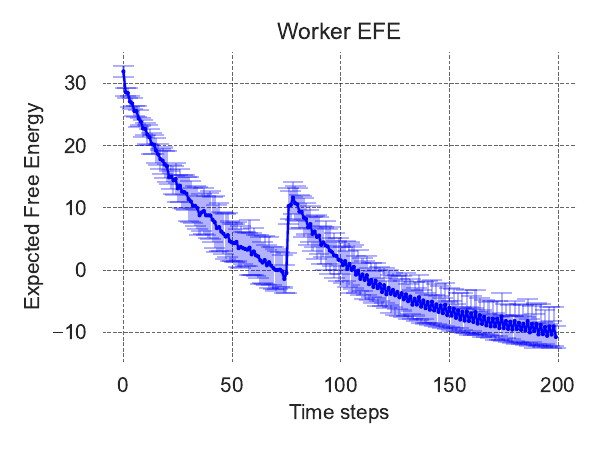}
    \captionsetup{skip=0pt}
    \caption{Expected Free Energy for the \w learning the transition model.}
    \label{fig:hw_change_l_efe_w}
\end{subfigure}
\hspace{2cm}
\begin{subfigure}[t]{.4\linewidth}
    \centering
    \includegraphics[width=0.75\linewidth]{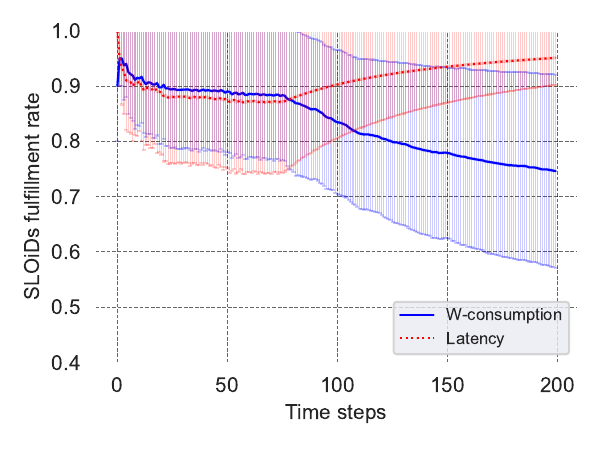}
    \captionsetup{skip=0pt}
    \caption{\w SLOiDs fulfillment with known transition model and facing a sudden environmental change.}
    \label{fig:env_change_nl}
\end{subfigure}
\hspace{1cm}
\begin{subfigure}[t]{.4\linewidth}
    \centering
    \includegraphics[width=0.75\linewidth]{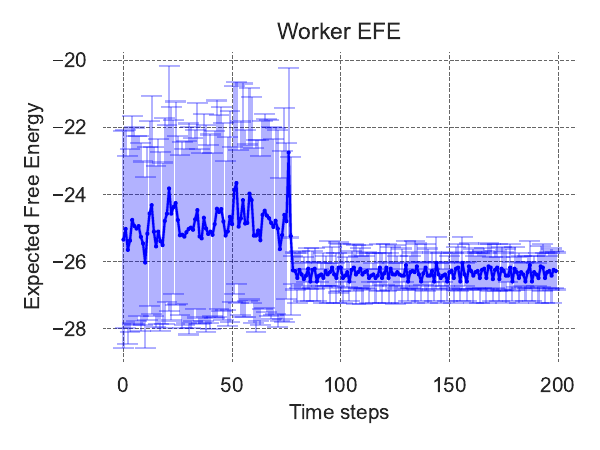}
    \captionsetup{skip=0pt}
     \caption{Expected Free Energy for the \w without learning the transition model.}
     \label{fig:hw_change_nl_efe_w}
\end{subfigure}
\vspace{6pt}
\caption{AIF \w agent suffering an offloading to a new and different host after 75 time steps controlling its SLOiDs.}
\label{fig:env_change}
\end{figure}

Figure \ref{fig:env_change} illustrates the experimental results when the hosting environment of the AIF agents was deliberately and suddenly altered, simulating an offloading of their services to different hardware.
Specifically, Figure \ref{fig:env_change_l} displays the performance of the AIF \w agent while learning the transition model, whereas Figure \ref{fig:env_change_nl} shows the same \w agent operating with a known transition dynamics model.
Prior to analyzing these results, it is crucial to note that the new host exhibits significantly higher energy consumption compared to the initial one.
Consequently, fulfilling the \slowc SLOiD becomes exceedingly challenging without redefining it.
Bearing this in mind, we observe a clear degradation in the \slowc SLOiD for both \w agents. This degradation is more pronounced for the learning agent.
However, the learning agent demonstrates a faster reduction in the variance of the \sloi SLOiD, which can be attributed to its superior adaptability.
Furthermore, Figure \ref{fig:hw_change_l_efe_w} presents the expected free energy (EFE) for the AIF \w learning agent during the same experiment.
It is evident that the agent initially reduces its EFE. However, the EFE suddenly increases due to the environmental change, after which the agent successfully resumes its EFE reduction.
In contrast, Figure \ref{fig:hw_change_nl_efe_w} depicts the EFE for the AIF \w non-learning agent. This agent also experiences a change in its EFE, primarily characterized by a substantial reduction in its variance. This suggests a narrowed range of possible actions for the agent following the offloading process.

\subsection{Computing cost}
\label{subsec:comp_cost}
Two main factors influence the computational cost for AIF agents: (1) learning the transition model, and (2) the length of the planned policy. In this analysis, computational cost primarily refers to execution time.

On average and in a Apple M3 chip with 18GB of RAM, a single time step in the environment takes approximately $80.55$ seconds when the transition model must be learned for a policy length of $3$. In contrast, when transitions are provided, the same step takes only $1.17$ seconds on average. While these durations naturally depend on the computational capacity of the device, the ratio between them offers a more device-agnostic metric: learning the transition model results in a slowdown of approximately $69.08\times$.
Interestingly, this computational burden diminishes with shorter policy horizons due to the reduced combinatorial complexity. When repeating the experiment with a policy length of $1$, the slowdown ratio is halved, the learning agent is approximately $35\times$ slower than its counterpart with provided transitions.

\begin{figure}[ht!]
  \centering
\begin{subfigure}[t]{.3\linewidth}
    \centering
    \includegraphics[width=0.9\linewidth]{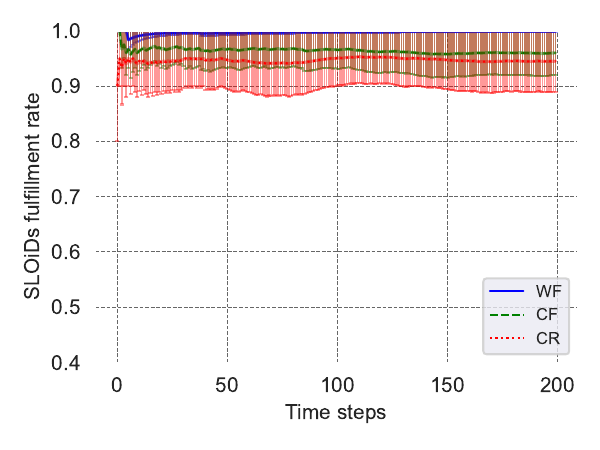}
    \captionsetup{skip=0pt}
    \caption{\p SLOiDs fulfillment rate, they are \slowd, \slocdf, and \slocdr.}
    \label{fig:l_pl1_slo_w}
\end{subfigure}
\hspace{0.25cm}
\begin{subfigure}[t]{.3\linewidth}
    \centering
    \includegraphics[width=0.9\linewidth]{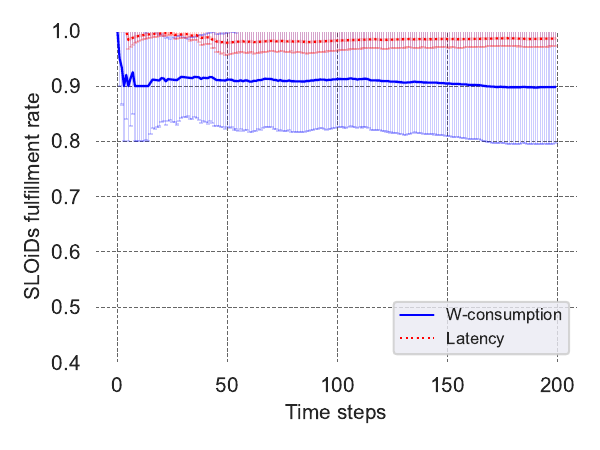}
    \captionsetup{skip=0pt}
     \caption{\w SLOiDs fulfillment rate, they are \sloi and \slowc.}
     \label{fig:l_pl1_slo_p}
\end{subfigure}
\hspace{0.25cm}
\begin{subfigure}[t]{.3\linewidth}
    \centering
    \includegraphics[width=0.9\linewidth]{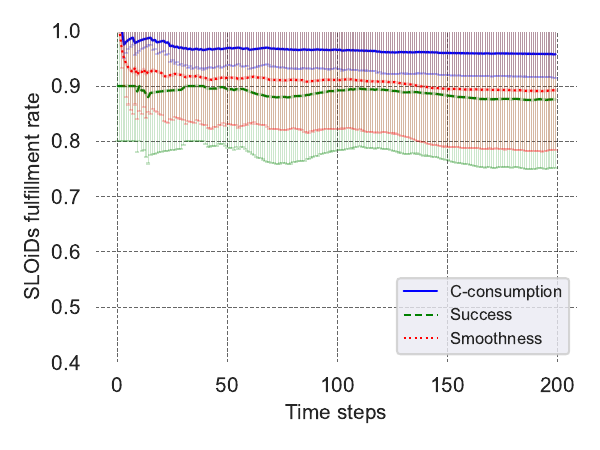}
    \captionsetup{skip=0pt}
    \caption{\co SLOiDs fulfillment rate, they are \slos, \slod, and \slocc.}
    \label{fig:l_pl1_slo_co}
\end{subfigure}
\caption{SLOiDs fulfillment rate with \textbf{B} parameter learning and policy length of 1.}
\label{fig:l_pl1_slos}
\end{figure}

Given this computational overhead, it is important to evaluate agent performance under constrained policy lengths. To this end, we assess SLOiD fulfillment rates under the same setup used in previous evaluations. When agents had to learn the transitions with a policy length of $1$, agents achieved fulfillment rates comparable to those with a policy length of $3$, see Figure \ref{fig:l_pl1_slos}. Surprisingly, the \p agent fulfillment rate (Figure \ref{fig:l_pl1_slo_p}) was consistently higher for length-$1$ policies than for length-$3$, which was unexpected.
This result may stem from the relatively small number of repetitions ($10$) and the stochastic nature of the environment, where identical actions in the same state can yield different outcomes. Additionally, most agent actions are boolean, and the environment exhibits limited temporal dependencies. As a reminder, environmental responses are based on the agent’s action in the previous state, and new states are sampled to fulfill constraints without necessarily maintaining temporal continuity.
Consequently, in this scenario, policy length may not be a critical factor in minimizing expected free energy (EFE)\footnote{Although longer policies generally achieve lower EFE values, this effect is partly an artifact of EFE computation: longer policies distribute cumulative EFE over more steps, which lowers the average per step.}. However, it is a key factor to consider in order to minimize the resource requirements of the agents.

\section{Discussion}
\label{sec:discussion}

The results indicate that AIF agents effectively support distributed intelligence in the CC, enabling interdependent services to collaborate while maintaining consistent performance, quality, and cost. This work represents a first step toward realizing that vision. In what follows, we examine the gap between our current implementation and the broader potential of intelligence distribution using AIF in the CC. We organize the discussion into design-time and runtime considerations, while acknowledging that several aspects span both.

\subsection{Design}

\subsubsection{AIF Agent Model}

This work introduces an AIF agent model for supervising services based on predefined SLOiDs. Defining these SLOiDs is critical to ensure service compliance with performance expectations in its execution environment. This task is simpler when prior deployment knowledge exists; otherwise, changes in host devices can compromise SLOiDs fulfillment due to mismatched assumptions.

Modeling the AIF agent requires identifying dependencies between SLOiD variables and system variables. These dependencies—especially those involving inter-service interactions may remain hidden without cross-service analysis~\cite{casamayor_pujol_deepslos_2024}. Causal discovery methods like those by Mariani et al.~\cite{mariani_distributed_2024} offer potential support, although typically post-deployment. In this study, we modeled service data as Bayesian Networks (BNs) and derived Markov Blankets as a prerequisite for both learned and predefined transition models~\cite{sedlak_equilibrium_2024}.

Once dependencies are identified, they must be quantified, leading to two key design choices: (i) whether to model variables as discrete or continuous, and (ii) how to acquire the relationship parameters.

We chose discrete variables for computational tractability. This required determining variable granularity, number of classes and class thresholds, effectively coarse-graining the agent's model and limiting behavioral resolution. Continuous modeling, while more expressive, demands richer data and assumptions on variable distributions~\cite{catal_learning_2020}, and is computationally intensive.

Regarding quantification, we explored both deterministic (expert-based) and learned approaches (see subsections \ref{res:slo_fulfill_rate} and \ref{res:slo_fulfill_rate_learning}). Expert-based models are stable but rigid, and increasingly inadequate as system complexity grows. Learned models offer adaptability and autonomy, but incur high computational cost and demand longer convergence times.

\subsubsection{Deployment}

Although our experiments relied on simulations, deploying AIF agents alongside services within a real CC infrastructure is essential. This requires tight integration between services and their supervisory agents to enable real-time SLOiD adaptation or parameter adjustments. Such integration promotes a novel design paradigm in which services natively support adaptive supervision through new interfaces, enabling autonomous control and continuous optimization.

\subsection{Runtime}

At runtime, a core challenge is the combinatorial explosion of state-action pairs. To mitigate this, we used binary actions (e.g., \actOff, \actOn), and occasionally a neutral option (\actSt), which reduced complexity but limited precision. This trade-off can hinder the agent's ability to reach target states efficiently when fine-grained control is needed. The effect worsens with longer policy lengths, making it critical to match action granularity and policy horizon to the environment’s dynamics.
We propose addressing this with hierarchical action-selection: AIF agents choose high-level actions, while a secondary controller handles precise execution~\cite{pezzato_novel_2020, parr_discrete_2018, tschantz_simulating_2022}. Such hybrid strategies, as suggested in~\cite{collis_learning_2024}, could improve adaptability and efficiency.

Flexibility is also vital in dynamic settings. Services may adjust preferences based on context or stakeholders. AIF agents accommodate this via direct updates to expected outcomes (C), adapting to runtime goals. Software updates, common in CC, may alter service behavior or capabilities. While minor behavior changes can be handled through continual learning, new capabilities may obsolete the agent's model. Detecting and adapting to such changes—or enabling agents to extend their models accordingly—remains an open challenge.
This leads to the concept of model transfer, sharing models across similar services or devices could accelerate adaptation and support federated learning-like knowledge aggregation. Future work should explore mechanisms for safe and effective model transfer between AIF agents.


\section{Related work}
\label{sec:rw}
In the context of this work, we identified two topics that have been addressed to some extent in existing work, namely adaptive mechanisms for the Computing Continuum, and distributed intelligence through Active Inference. The first topic targets the problem domain and the latter one the methodology applied. We present a concise description of related work and contrast it with our presented work to highlight the research gap we aim to fill.

\subsection{Adaptive Mechanisms for the Computing Continuum}

The CC is an emerging paradigm that has only been developed and extended over the last recent years of research~\cite{dustdar_distributed_2023}; in particular, there remain multiple challenges in orchestrating such large-scale systems~\cite{casamayor_pujol_fundamental_2023}, which require intelligent adaptation mechanisms. 
As such, Filinis et al.~\cite{filinis_intent-driven_2024} present an intelligent auto-scaling agent for the CC, which scales the replicas of serverless functions that are deployed over a continuum.
Closely related to that, Zafeiropoulos et al.~\cite{zafeiropoulos_ai-assisted_2024} provide a Reinforcement Learning (RL)-based auto-scaling environment for CC systems that explored the synergies between low- and high-level controlling entities to achieve globally optimal solutions.
Their core motivation was to combine SLOs with dynamic processing requirements, which also applies to \textit{Octopus}, a framework developed by Zhang et al.~\cite{zhang_octopus_2023}. \textit{Octopus} finds optimal service configurations in multi-tenant edge computing scenarios; for this, it predicts SLO fulfillment of two variables based on a deep neural network. To detect SLO violations of a scheduling task, Shubha et al.~\cite{shubha_adainf_2023} presented \textit{AdaInf}, which can find SLO-fulfilling resource allocations between model training and inference; such mechanisms will prove essential to the CC.

While research on adaptive cloud mechanisms has made significant progress~\cite{verma_auto-scaling_2021}, the CC is still unexploited, but recent advances in Edge orchestration techniques promise to improve this. For example, through offloading techniques between mobile Edge devices and stationary computing resources, like Fog or Cloud. As such, Ma et al.~\cite{ma_video_2024}, Wu et al.~\cite{wu_intelligent_2023}, and Spring et al.~\cite{spring_mach_2025} provide orchestration mechanisms that can be integrated into the CC. To make accurate predictions of how intent-based services in the CC will behave, Akbari et al. presented \textit{iContinuum}~\cite{akbari_icontinuum_2024}, a simulation environment that can run experiments in Edge-to-Cloud scenarios.

Given the presented research, we conclude that research on the CC is still at its beginning, if not at an infant stage, with most works focusing on the potentials and challenges of the CC and comparably few empirically-evaluated solutions. As Edge computing advances, including the research on multiple distributed agents in Edge networks, orchestration mechanisms, as presented in this paper, will be heavily needed to ensure collaboration between multiple computational tiers.

\subsection{Distributed Intelligence through Active Inference}

While AIF as a concept has been developed and enhanced over recent years \cite{friston2024supervised,friston2023active}, its transition towards applications is still at a relatively early stage \cite{friston_designing_2024}. Outside of neuroscience, AIF was mostly applied to robotics, as shown by the work of Lanillos et al.~ \cite{lanillos_active_2021} and Oliver at al.~\cite{oliver_empirical_2022}, who use it as a framework for sensing and perception. In particular, Oliver et al. give a comprehensive overview of how AIF allows (robotic) systems to act under uncertainty. Complementarily, De Vries et al.~\cite{de_vries_toward_2023} formulate a general design for AIF agents that applies across disciplines.

Nevertheless, applications of AIF to continuous stream processing systems are still scarce, with some exception, such as the works of Sedlak et al.~\cite{sedlak_equilibrium_2024,sedlak_adaptive_2024}. There, the authors show how to ensure SLOs of continuous video processing tasks by training a service interpretation model solely from observations; models were then shared between agents to speed up convergence to globally optimal solutions. In another work, Danilenka et al.~\cite{danilenka_adaptive_2024} developed an AIF agent that could optimize the SLO fulfillment of a federated learning task. Given the heterogeneity within the CC, they used AIF to find the optimal training configuration for individual clients. Levchuk et al.~\cite{levchuk_active_2019} provide another example of a multi-agent system in which AIF is used to guide the exploration-exploitation tradeoff of multiple distributed agents. Noteworthy, they showed how agents' perception and collaboration could improve the convergence of sensing tasks.

Given the presented work, we conclude that AIF is regarded as a very promising solution for agent-based sensing environments, as can be found natively in the IoT domain. While some concepts of AIF have made their transition to machine learning and reinformcement learning \cite{mazzaglia2022free,tschantz2020scaling, tschantz2020reinforcement}, wholesome solutions that encompass an entire software system are still scarce. To that extent, the research presented in this paper provides a well-needed guideline on how to design AIF agents in distributed computing scenarios. AIF can be of particular advantage in large-scale computing networks, like the CC, where empirically verifiable solutions are needed to prove the correctness of an approach.

\section{Conclusions}
\label{sec:conclusions}
In this article, we explored the use of AIF to distribute intelligence across a multi-service application within the CC. Specifically, we focused on a video stream processing pipeline requiring face blurring before delivery to the final consumer. This was achieved by deploying an AIF agent for each service and equipping them with mechanisms for elastic adaptation to service-specific requirements.

Our findings suggest that AIF is a promising approach for managing distributed applications in the CC. The AIF agents effectively fulfilled each service’s SLOiDs by selecting targeted actions and, when necessary, adapting based on the outcomes of their decisions. When provided with expert-defined transition models, the agents achieved over 90\% SLOiD fulfillment rates. In contrast, agents that learned their models through interaction with the environment achieved rates approximately 10\% lower.
When compared to state-of-the-art alternatives such as multi-agent reinforcement learning, we observed that extensively trained MARL agents slightly outperformed AIF agents. However, under minimal training conditions, AIF agents exhibited superior performance.
We also analyzed agent behavior during service offloading scenarios, where the deployment environment changed. The AIF agents demonstrated adaptability by detecting the environmental shift and exploring new policies to recover fulfillment rates. Notably, agents learning their own transition models were capable of strategically deprioritizing unachievable SLOiDs in favor of maximizing achievable ones, showcasing a high degree of flexibility.
Finally, we evaluated the computational cost of AIF agents in terms of execution time. A key factor in their efficiency is the selection of an appropriate policy length (i.e., look-ahead depth), as longer horizons can significantly increase computation time, especially in complex environments.

Looking ahead, AIF agents can be developed through a combination of expert knowledge, where the consequences of actions are predefined, and autonomous lifelong learning, where agents learn action outcomes over time. This hybrid design strategy supports the adaptability needed to address the heterogeneity and dynamism inherent in CC environments.
We have also identified several gaps and promising directions for future research aimed at further aligning AIF capabilities with the unique demands of the CC. From a design perspective, further work is needed to enhance model flexibility, improve action predictability, and integrate more complex state-action relationships that better reflect real-world CC scenarios. On the runtime side, tackling challenges such as the action-state explosion and adapting models in response to environmental changes is critical for deploying robust and accountable AIF-driven services.
Some of these challenges have been addressed in the broader AIF literature and could be adapted for CC applications. For instance, alternative planning algorithms have been proposed to enable faster \cite{champion_deconstructing_2023, paul2024efficient} and more sophisticated \cite{friston2021sophisticated, friston2023intentional} action selection. Another promising direction is the incorporation of partially observed environments with a form of theory of mind, enabling agents to infer the internal states, intentions, and likely future actions of other agents \cite{albarracin2024shared}. Combining these avenues will be crucial for modeling and managing the complex interactions that arise among large numbers of agents in distributed systems.

In summary, this study demonstrates that Active Inference offers a powerful foundation for distributed intelligence in the Computing Continuum. AIF agents can not only fulfill predefined service-level objectives but also adapt autonomously to evolving conditions. These findings lay the groundwork for future advancements in AIF-driven service orchestration and highlight the potential of AIF as a tool for enabling resilient, intelligent, and autonomous service management across the CC.

\section*{Acknowledgments}
This work is supported by CNS2023-144359 financed by MICIU/AEI/10.13039/501100011033 and the European Union NextGenerationEU/PRTR.\\
Thanks to Dimitrije Markovic for his support.


\appendix
\section*{Appendix}
\appendix
\label{sec:appendix}
\section{\p's transition model}
\label{sec:appendix-producer}

\paragraph{\slowd Modality}

The transition probabilities for the \slowd\ modality specify the distribution $P(\slowd_{t+1} | \slowd_t, \textit{FPS}_t, \textit{Change\_FPS}_t)$. These probabilities depend on the current state of the modality ($\slowd_t$), the current \textit{FPS} state ($\textit{FPS}_t$), and the relevant action component ($\textit{Change\_FPS}_t$). Fully specifying the corresponding Conditional Probability Table (CPT) explicitly would require defining a 3x3 transition matrix for each of the 5 \textit{FPS} values and 3 \textit{Change\_FPS} values (15 matrices total). However, we define the CPT completely and more compactly using the following rules:

\begin{enumerate}
    \item If $\textit{Change\_FPS}_t = \actSt$, the state remains unchanged: $\slowd_{t+1} = \slowd_t$.
    \item The state also remains unchanged if boundary conditions for \textit{FPS} prevent the action's effect: i.e., if $\textit{Change\_FPS}_t = \actInc$ and $\textit{FPS}_t$ is already the highest value, or if $\textit{Change\_FPS}_t = \actDec$ and $\textit{FPS}_t$ is already the lowest value.
    \item If $\textit{Change\_FPS}_t = \actInc$ (and \textit{FPS} is not maximum):
        \begin{itemize}
            \item If $\slowd_t = \actDec$ or $\slowd_t = \actSt$, then $\slowd_{t+1} = \actDec$.
            \item If $\slowd_t = \actInc$, then $\slowd_{t+1} = \actSt$.
        \end{itemize}
    \item If $\textit{Change\_FPS}_t = \actDec$ (and \textit{FPS} is not minimum):
        \begin{itemize}
            \item If $\slowd_t = \actInc$ or $\slowd_t = \actSt$, then $\slowd_{t+1} = \actInc$.
            \item If $\slowd_t = \actDec$, then $\slowd_{t+1} = \actSt$.
        \end{itemize}
\end{enumerate}
Note that these rules simplify the dependency on the \textit{FPS} state, using it only to check boundary conditions. The precise dynamics might vary with specific hardware configurations in practice, but this expert model captures the intended behaviour.

\paragraph{\slocdf \& \slocdr Modalities}
The transition dynamics for \slocdf are defined analogously, depending on itself at $t$, $\textit{FPS}_t$, and $\textit{Change\_FPS}_t$. The dynamics for \slocdr follow the same pattern but depend on $\slocdr_t$, $\textit{Resolution}_t$, and $\textit{Change\_Resolution}_t$.

\paragraph{FPS \& Resolution Modailities}
The transitions for the \textit{FPS} and \textit{Resolution} modalities themselves are fully deterministic. These dynamics are defined by $P(\textit{FPS}_{t+1} | \textit{FPS}_t, \textit{Change\_FPS}_t)$ and $P(\textit{Resolution}_{t+1} | \textit{Resolution}_t, \textit{Change\_Resolution}_t)$, respectively. Their dependencies are shown in Figure~\ref{fig:producer_schema}, and their CPTs are defined by the following rules:
\begin{enumerate}
    \item If the action (\textit{Change\_FPS} or \textit{Change\_Resolution}) is \actSt, the value remains unchanged at $t+1$.
    \item If the action is \actInc, the value at $t+1$ transitions to the next higher discrete level defined for the modality, unless the value at $t$ is already the maximum, in which case it remains unchanged.
    \item If the action is \actDec, the value at $t+1$ transitions to the next lower discrete level, unless the value at $t$ is already the minimum, in which case it remains unchanged.
\end{enumerate}

\section{\w's transition model}
\label{sec:appendix-worker}

\paragraph{\sloi~Modality}
The transition $P(\sloi_{t+1} | \sloi_t, \textit{Toggle\_Comm}_t)$ is simplified to depend only on the current state $\sloi_t$ and the action taken on the upstream communication channel ($\textit{Toggle\_Comm}_t$)\footnote{\sloi also depends on the \textit{Execution Time} and the \textit{FPS}, however, here we focus on the communication action to simplify all definitions.}. 
The CPT is defined by the following deterministic rules:
\begin{enumerate}
    \item If $\textit{Toggle\_Comm}_t = \actEn$, the state transitions to or remains \textit{True}: $\sloi_{t+1} = \textit{True}$.
    \item If $\textit{Toggle\_Comm}_t = \actDis$, the state remains unchanged: $\sloi_{t+1} = \sloi_t$.
\end{enumerate}
This simplification assumes controlling the communication channel fully dictates this SLO status in our model. All state modalities affected by the action \textit{Toggle\_Comm} assume that the \p agent will support the request sent by the other agents.

\paragraph{Execution Time Modality}
The transition $P(\textit{ExecTime}_{t+1} | \textit{ExecTime}_t, \textit{GPU}_t, \textit{Switch\_GPU}_t)$ depends on the current \textit{Execution Time}, the current \textit{GPU} state, and the action applied to the GPU. The CPT is defined by these rules:
\begin{enumerate}
    \item If $\textit{Switch\_GPU}_t = \actSt$, or if $\textit{Switch\_GPU}_t = \actOff$ when $\textit{GPU}_t = \textit{Off}$, or if $\textit{Switch\_GPU}_t = \actOn$ when $\textit{GPU}_t = \textit{On}$. Then, the \textit{Execution Time} remains unchanged ($\textit{ExecTime}_{t+1} = \textit{ExecTime}_t$).
    \item If the GPU is on (i.e., $\textit{Switch\_GPU}_t = \actOn$ and $\textit{GPU}_t = \textit{Off}$, leading to $\textit{GPU}_{t+1} = \textit{On}$), the \textit{Execution Time} decreases by one discrete step ($\textit{ExecTime}_{t+1} = \text{prev}(\textit{ExecTime}_t)$). If $\textit{ExecTime}_t$ is already the lowest value, it remains unchanged.
    \item If the GPU is off (i.e., $\textit{Switch\_GPU}_t = \actOff$ and $\textit{GPU}_t = \textit{On}$, leading to $\textit{GPU}_{t+1} = \textit{Off}$), the \textit{Execution Time} increases by one discrete step ($\textit{ExecTime}_{t+1} = \text{next}(\textit{ExecTime}_t)$). If $\textit{ExecTime}_t$ is already the highest value, it remains unchanged.
\end{enumerate}

\paragraph{FPS Modality}
The transition $P(\textit{FPS}_{t+1} | \textit{FPS}_t, \textit{Toggle\_Comm}_t)$ depends on the current FPS value and the action on the upstream communication channel.
\begin{enumerate}
    \item If communication is disabled or unchanged ($\textit{Toggle\_Comm}_t = (\actDis \text{ or } \actSt$)), the FPS remains the same: $\textit{FPS}_{t+1} = \textit{FPS}_t$. 
    \item If communication is enabled ($\textit{Toggle\_Comm}_t = \actEn$), the \textit{FPS} decreases by one discrete step: $\textit{FPS}_{t+1} = \text{prev}(\textit{FPS}_t)$. If $\textit{FPS}_t$ is already the lowest value, it remains unchanged.
\end{enumerate}

\paragraph{\slowc~Modality}
The $P(\slowc_{t+1} | \slowc_t, \textit{ShareInfo}_t, \textit{GPU}_t, \textit{Toggle\_Comm}_t, \textit{Switch\_GPU}_t)$ depends on the current \slowc state, the states of \textit{Share Information} and \textit{GPU}, and the actions applied to them. We assume the resulting \slowc state depends on whether the \textit{ShareInfo} and \textit{GPU} states are \textit{effectively activated or deactivated} by the actions taken. The CPT rules are:
\begin{enumerate}
    \item If both \textit{Share Information} and \textit{GPU} are activated (transition to or stay \textit{True}), \slowc increases by one step: $\slowc_{t+1} = \text{next}(\slowc_t)$.
    \item If one modality (\textit{ShareInfo} or \textit{GPU}) is activated while the other's state remains unchanged, \slowc increases by one step: $\slowc_{t+1} = \text{next}(\slowc_t)$.
    \item If either modality is deactivated (transitions to or stays \textit{False}), \slowc decreases by one step: $\slowc_{t+1} = \text{prev}(\slowc_t)$. 
    \item If one modality is activated and the other is deactivated within the same time step, \slowc remains the same: $\slowc_{t+1} = \slowc_t$. 
    \item If both modalities' states remain unchanged, \slowc remains the same: $\slowc_{t+1} = \slowc_t$. 
\end{enumerate}

\paragraph{Share Info Modality}
The transition $P(\textit{ShareInfo}_{t+1} | \textit{ShareInfo}_t, \textit{Toggle\_Comm}_t)$ depends on the current state and the action on the upstream communication channel. The CPT rules are:
\begin{enumerate}
    \item If the action is \actDis, the state becomes False: $\textit{ShareInfo}_{t+1} = \textit{False}$.
    \item If the action is \actEn, the state becomes True: $\textit{ShareInfo}_{t+1} = \textit{True}$.
    \item If the action is \actSt, the state remains unchanged: $\textit{ShareInfo}_{t+1} = \textit{ShareInfo}_t$. 
\end{enumerate}

\paragraph{GPU Modality}
The transition $P(\textit{GPU}_{t+1} | \textit{GPU}_t, \textit{Switch\_GPU}_t)$ depends on the current state and the action applied to the GPU. The CPT rules mirror those for \textit{Share Info}, using actions \actOn, \actOff, and \actSt:
\begin{enumerate}
    \item If $\textit{Switch\_GPU}_t = \actOff$, then $\textit{GPU}_{t+1} = \textit{False}$.
    \item If $\textit{Switch\_GPU}_t = \actOn$, then $\textit{GPU}_{t+1} = \textit{True}$.
    \item If $\textit{Switch\_GPU}_t = \actSt$, then $\textit{GPU}_{t+1} = \textit{GPU}_t$.
\end{enumerate}

\section{\co's transition model}
\label{sec:appendix-consumer}

\paragraph{\slos~Modality}
The transition $P(\slos_{t+1} | \slos_t, \textit{Resolution}_t, \textit{Toggle\_Comm}_t)$ depends on the current \slos state, the \textit{Resolution}, and the upstream communication action. The CPT is defined by the following rules, which currently simplify the dependency on $\textit{Resolution}_t$: 
\begin{enumerate}
    \item If $\textit{Toggle\_Comm}_t$ is \actEn, the state transitions to or remains \textit{True}: $\slos_{t+1} = \textit{True}$.
    \item If $\textit{Toggle\_Comm}_t$ is \actDis or \actSt, the state remains unchanged: $\slos_{t+1} = \slos_t$.
\end{enumerate}

\paragraph{\slod~Modality}
The transition $P(\slod_{t+1} | \slod_t, \textit{FPS}_t, \textit{Toggle\_Comm}_t)$ depends on the current \slod state, the \textit{FPS} state, and the upstream communication action. The CPT rules are:
\begin{enumerate}
    \item If $\textit{Toggle\_Comm}_t$ is \actDis or \actSt, the state remains unchanged: $\slod_{t+1} = \slod_t$.
    \item If $\textit{Toggle\_Comm}_t$ is \actEn: the resulting state $\slod_{t+1}$ becomes \textit{True}, reflecting the expectation that enabling communication aims to improve this SLO, potentially by facilitating an FPS increase. 
\end{enumerate}

\paragraph{\slocc~Modality}
The transition $P(\slocc_{t+1} | \slocc_t, \textit{ShareInfo}_t, \textit{Toggle\_Comm}_t)$ depends on the current \slocc state, the current \textit{Share Information} state, and the upstream communication action. The CPT rules are:
\begin{enumerate}
    \item If $\textit{Toggle\_Comm}_t = \actEn$ causes \textit{Share Information} to switch from \textit{False} to \textit{True} (e.g., $\textit{ShareInfo}_t = \textit{False}$), then \slocc increases: $\slocc_{t+1} = \text{next}(\slocc_t)$.
    \item If the action $\textit{Toggle\_Comm}_t = \actDis$ causes \textit{Share Information} to switch from \textit{True} to \textit{False} (e.g., $\textit{ShareInfo}_t = \textit{True}$), then \slocc decreases: $\slocc_{t+1} = \text{prev}(\slocc_t)$.
    \item In all other cases, \slocc remains unchanged: $\slocc_{t+1} = \slocc_t$.
\end{enumerate}

\paragraph{FPS and Resolution Modalities}
The transitions $P(\textit{FPS}_{t+1} | \textit{FPS}_t, \textit{Toggle\_Comm}_t)$ and $P(\textit{Resolution}_{t+1} | \textit{Resolution}_t, \textit{Toggle\_Comm}_t)$ depend on their respective current states and the upstream communication action. The deterministic CPT rules are:
\begin{enumerate}
    \item If $\textit{Toggle\_Comm}_t$ is \actDis or \actSt, the respective modality's value remains unchanged ($\textit{FPS}_{t+1} = \textit{FPS}_t$, $\textit{Resolution}_{t+1} = \textit{Resolution}_t$).
    \item If $\textit{Toggle\_Comm}_t$ is \actEn, both the \textit{FPS} and \textit{Resolution} values increase by one discrete step, if possible (i.e., $\textit{FPS}_{t+1} = \text{next}(\textit{FPS}_t)$, $\textit{Resolution}_{t+1} = \text{next}(\textit{Resolution}_t)$), subject to saturation at their maximum values.
\end{enumerate}

\paragraph{Share Information Modality}
The transition probabilities $P(\textit{ShareInfo}_{t+1} | \textit{ShareInfo}_t, \textit{Toggle\_Comm}_t)$ are defined identically to the corresponding modality in the \w\ agent model. 

\end{document}